\documentclass[amsmath, amsfonts, amssymb, aps, superscriptaddress]{revtex4}

\usepackage{tabularx}
\usepackage{graphicx}
\usepackage[font=scriptsize]{caption} 
\usepackage{float}
\usepackage{dcolumn}
\usepackage{bm}
\usepackage{mathrsfs}
\usepackage{amsmath}

\newcommand{\be}{\begin{eqnarray}}
\newcommand{\ee}{\end{eqnarray}}
\newcommand{\om}{\alpha}
\usepackage{amsfonts} 

\usepackage{color}

\definecolor{darkred}{rgb}{.8,0,0}

\definecolor{darkblue}{rgb}{0,0,.7}

\definecolor{darkgreen}{rgb}{0,.7,0}

\begin{document}

\title{Spectral Analysis of Quasinormal Modes of Planck Stars}

\author{Davide Batic}
\email{davide.batic@ku.ac.ae}
\affiliation{Mathematics Department, Khalifa University of Science and Technology, PO Box 127788, Abu Dhabi, United Arab Emirates}

\author{Denys Dutykh}
\email{denys.dutykh@ku.ac.ae}
\affiliation{Mathematics Department, Khalifa University of Science and Technology, PO Box 127788, Abu Dhabi, United Arab Emirates}

\author{Fabio Scardigli}
\email{fabio@phys.ntu.edu.tw}
\affiliation{Dipartimento di Matematica, Politecnico di Milano, Piazza Leonardo da Vinci 32, 20133 Milano, Italy}
\affiliation{Department of Applied Mathematics, University of Waterloo, Ontario N2L 3G1, Canada}

\date{\today}

\begin{abstract}
We investigate the quasinormal modes (QNMs) of Planck stars within the framework of scale-dependent gravity (SDG). In our setup, the running parameter $\alpha$ is fixed to a negative value by matching the effective Newtonian potential to the one-loop EFT result. As a consequence, the associated running Newton coupling does not realise the ultraviolet fixed point of asymptotically safe gravity, and the geometry should be interpreted as an SDG-inspired effective metric rather than a realisation of asymptotically safe gravity itself. We focus on the resulting renormalisation-group-improved Schwarzschild metric, which naturally yields a finite-size Planck-density core. Building on this background, we compute the QNM spectrum for scalar, electromagnetic, and gravitational perturbations using the Spectral Method (SM). This approach, known for its superior accuracy over high-order WKB schemes, enables the detection of fundamental modes, large families of overtones, and purely imaginary overdamped modes that are entirely missed in previous analysis. Our results reveal a robust \emph{Martini glass} morphology of the oscillatory spectrum across perturbation sectors, nearly equally spaced overdamped modes with characteristic anomalous gaps, and the emergence, in the gravitational sector, of isolated overdamped modes separated from the main sequence by exceptionally large frequency intervals. These features, resolved here for the first time in the Planck-star context, underscore the importance of high-precision spectral techniques in probing subtle signatures of quantum-gravity–inspired black hole models.      
\end{abstract}

\maketitle

\section{Introduction}\label{Intro}

The quest for a complete theory of gravity in ultraviolet (UV) space remains a central challenge in contemporary theoretical physics. The straightforward quantisation of general relativity (GR) leads to a perturbatively non-renormalisable quantum field theory \cite{Sagnotti1985}, requiring, in principle, the adjustment of infinitely many counterterms to make physical predictions: a situation that renders this approach practically unusable. It has long been known that adding higher-derivative terms to the gravitational action can yield a perturbatively renormalizable theory \cite{Stelle1977}. However, such terms typically introduce ghost degrees of freedom, which can lead to instabilities. An alternative route, first proposed by Weinberg \cite{Weinberg1, Weinberg2} and developed further by others (see, e.g., \cite{Wetterich93, Reuter98, Bonanno2000PRD, Niedermaier2006}), is that a quantum theory of gravity might \textit{dynamically} evade the divergences of perturbative gravity. In this framework, known as \textit{asymptotically safe gravity} (ASG), the central idea is the existence of a non-Gaussian fixed point of the gravitational renormalisation group (RG) flow, which governs the high-energy behaviour of the theory and ensures the absence of unphysical UV divergences. The RG flow can be integrated to obtain a running Newton’s constant $G(k)$ as a function of the momentum scale $k$.

In the context of black hole physics, this running coupling $G(k)$ can be incorporated into the classical solution to produce an improved lapse function. At this stage, however, $G$ still depends on an arbitrary renormalisation scale $k$, and one must identify a physically motivated relation between $k$ and the radial coordinate $r$ of the spacetime. Once this identification is made, the composite function $G(k(r))$ can be reinserted into the classical metric, yielding a fully specified, quantum-improved lapse function. In practice, this identification is phenomenological and not unique. Different prescriptions for the map $k(r)$, all consistent with basic physical requirements, lead to different RG-improved geometries. In the present work, we adopt the scale-setting used in \cite{Bonanno2000PRD, Bonanno2004}, which defines a particular SDG realisation of the Planck-star scenario. Accordingly, our QNM spectra should be understood as predictions for this specific choice of $k(r)$ rather than as universal, scale-setting, independent features. Such extensions, motivated by the asymptotic safety program, naturally modify classical black hole geometries and encode quantum-gravity effects. ASG is one among several approaches in the literature (see, e.g., \cite{Jacobson95, Connes96, Ashtekar2005, Horava2009, Verlinde2011}) in which the fundamental couplings in the action, including Newton’s constant, the electromagnetic coupling, and the cosmological constant, acquire a scale dependence. Running couplings at the level of the effective action are a generic feature of quantum field theory. In the gravitational setting, such frameworks are broadly referred to as SDG theories. The scale dependence is expected to influence essential black hole properties, including the horizon structure, thermodynamics, and QNM  spectra, as explored, for example, in \cite{Koch16, Ricon19}. In the present work, we do not attempt to construct a fully asymptotically safe model of gravity or to provide a microscopic description of the matter sector sourcing the RG-improved geometry. We adopt an RG–improved Schwarzschild metric whose running Newton coupling has the same functional form as in ASG-inspired studies, but we fix its free parameter $\alpha$ by matching the effective Newtonian potential to the quantum-corrected result of GR viewed as an effective field theory. This procedure selects a negative value $\alpha < 0$, and thus a running Newton coupling that does not realise the positive ultraviolet fixed point of asymptotically safe gravity (see Section II). The resulting Planck-star spacetime should therefore be regarded as a model-dependent SDG effective background encoding quantum corrections, rather than as a prediction of asymptotically safe gravity in the strict sense. Correspondingly, we interpret the right-hand side of Einstein's equations as an effective stress–energy tensor $T_{\mu\nu}^{(eff)}$ encoding short-distance quantum-gravity effects. As in other quantum-corrected black-hole spacetimes, $T_{\mu\nu}^{(eff)}$ does not obey the standard classical energy conditions in the vicinity of the Planckian core and should not be regarded as arising from a conventional matter field with a well-defined equation of state.

The direct detection of gravitational waves from black hole mergers \cite{Abbott2016PRL, Abbott2019} has renewed interest in black hole perturbations \cite{Regge1957PR, Teukolsky1972} and their associated QNMs. These complex-frequency oscillations dominate the \emph{ringdown} phase, during which the merger remnant relaxes toward equilibrium via damped spacetime oscillations and gravitational-wave emission \cite{Ferrari2008}. QNMs depend only on the background geometry and the perturbation type (scalar, electromagnetic, gravitational, or fermionic) \cite{Cardoso2003, Rincon2020}, providing a unique spectral fingerprint of the black hole. Extensive studies have linked QNMs to black hole stability, gravitational-wave detection, and gauge/gravity dualities (see \cite{Berti2009CQG, Konoplya2011RMP} for reviews). For large astrophysical black holes, measured QNMs match those of the Schwarzschild solution to high precision, making quantum-gravity corrections such as those predicted by ASG effectively undetectable \cite{Liu2012}. In contrast, for miniature or primordial black holes, quantum effects can be significant, and QNMs offer a sensitive probe for distinguishing between competing regular black hole models. Compared to other observational tools, QNM spectroscopy is exceptionally precise \cite{Franchini2023}. While black hole shadow measurements with the Event Horizon Telescope carry uncertainties of order $10\%$, the fundamental QNM alone can constrain parameters at the $\gtrsim 90\%$ level. Next-generation detectors, coupled with the inclusion of overtone modes \cite{Spina24, Giesler25}, promise even greater accuracy, making QNMs a primary avenue for testing deviations from classical general relativity.

A substantial literature exists on QNMs of regular black hole models \cite{KonoplyaPLB2020, KonoplyaJCAP2022, KonoplyaPRD2023, Zinhailo2023}, including specific studies of the Hayward spacetime \cite{Flachi2013, Hendi2020, DuttaRoy2022} and the Bonanno-Reuter black hole \cite{Rincon2020}, as well as an approximate truncated form of the latter \cite{Liu2012}. However, several of the most significant and intriguing features of the QNM spectrum in these models have been overlooked. In particular, the behaviour of \emph{overtone} modes has received little attention. While it is often assumed that the fundamental mode dominates the gravitational-wave signal, numerical relativity simulations \cite{Giesler2019PRX} show that reproducing the full ringdown phase requires the inclusion of up to ten overtones. This finding not only underscores the physical relevance of overtones but also indicates that quasinormal ringing begins significantly earlier in the post-merger signal than previously expected.

Addressing this gap demands a computational approach capable of resolving both large families of overtones and other subtle spectral features with high precision. The SM is ideally suited for this purpose. Its advantages are well established \cite{Mamani2022EPJC, Batic2024EPJC, Batic2024CQG, Batic2024PRD, Batic2025EPJC, Batic2025CQG, Batic2025PRSA}: SM reproduces benchmark Schwarzschild QNMs with remarkable accuracy over a broad range of spins and multipoles, uncovers overdamped modes systematically missed by high-order WKB schemes, resolves extensive overtone spectra with spectral convergence, and remains numerically stable even in extreme parameter regimes. This combination of accuracy, completeness, and robustness makes SM particularly well-suited to the Planck-star scenario, where detecting isolated overdamped modes, anomalous intermode gaps, and other fine spectral structures requires exceptional resolution in the complex-frequency plane. Throughout this work, we use the term Planck star in a purely geometric sense. It denotes a black-hole spacetime in which the RG-improved metric develops a finite-size core of Planckian effective density enclosed by the event horizon. As discussed in Section III, the spacetime is not regular in the strict sense, since curvature invariants diverge at a finite radius inside the horizon, and our static model does not implement a fully non-singular quantum bounce. Accordingly, the background metric should be viewed as an SDG-motivated, model-dependent effective geometry, rather than as a generic or fundamental realisation of Planck-star physics.

A direct comparison with previous studies further underscores the need for our approach. In particular, \cite{Lambiase2023EPJC} does not verify that its results correctly reproduce the Schwarzschild QNM spectrum in the large-mass limit $M \gg 1$, a basic consistency check that is explicitly satisfied in our analysis. Moreover, their computation is restricted to the fundamental mode for scalar and vector perturbations, reported with relatively poor numerical accuracy, and omits entirely the scalar monopole case ($\ell = 0$), which we include here in full. Most critically, as is typical when relying on high-order WKB methods, \cite{Lambiase2023EPJC} fails to detect the purely imaginary, overdamped modes that emerge naturally in our spectral analysis. For comparison, \cite{KonoplyaJCAP2022} and \cite{Rincon2020} also rely on WKB-based schemes and similarly miss such modes, as well as the fine spectral structures revealed in the present work. Similar limitations are also observed in other recent works on regular or ASG-inspired black holes, such as \cite{Malik2024EPL, Stashko2024PRD}. These studies either fail to verify the large-mass Schwarzschild limit, restrict their analysis to a handful of low-lying modes, or rely on semi-analytic schemes that overlook the overdamped sector. This is a recurrent pattern in the WKB and analytical literature. While these methods can capture the fundamental mode in simple settings, they fail to reveal the full spectral structure, especially in scenarios such as the Planck-star case, where quantum corrections strongly influence the high-overtone and purely imaginary modes. By contrast, the SM employed here not only resolves the fundamental and higher-overtone modes with high precision but also identifies the overdamped sector, even in parameter regimes where WKB approximations produce incomplete or spurious spectra.

The paper is organised as follows. In Sections II and III, we review the SDG metric describing Planck stars, originally motivated by RG-improvement in the ASG program but here considered in the parameter regime $\alpha<0$, in which the running Newton coupling is no longer asymptotically safe, and summarise its key geometrical properties. Section IV introduces the perturbation equations for different spin fields in this background. In  Section V, we outline the implementation of the SM for computing QNMs, emphasising its numerical stability and accuracy. The resulting QNM spectra are presented and analysed in Section VI, with special attention to high overtones, overdamped modes, and the comparison with WKB-based results in the literature. Finally, Section VII summarises our findings and discusses their implications for the detectability of Planck-star signatures in gravitational-wave observations.

\section{Asymptotically safe or Scale-dependent gravity}

SDG provides a broad framework in which classical gravitational couplings are promoted to scale-dependent quantities. Specific examples, particularly black hole solutions, can be found in \cite{Koch16, Contreras2017, Contreras2018}. Within the ASG program, renormalisation-group--improved solutions in Newtonian gravity or general relativity are obtained by replacing the constant Newton coupling $G_N$ with a running coupling $G(k)$ and identifying a physically motivated relation $k(r)$ between the RG scale and the spacetime coordinate \cite{Bonanno2000PRD, Lambiase2022PRD}. In the formulation adopted here,
\be
\label{Gk}
G(k)= \frac{G_N}{1+ \om G_N k^2/\hbar},
\qquad {\rm with} \qquad k(r) = \hbar\left(\frac{r+\gamma G_N M}{r^3} \right)^{1/2},
\ee
where $\om$ and $\gamma$ are dimensionless constants, $c = k_B = 1$, and $\hbar$ is retained explicitly. This scale-setting prescription is phenomenological. It is motivated by the analyses of \cite{Bonanno2000PRD, Bonanno2004} and by the requirement of reproducing the quantum-corrected Newtonian potential once $\alpha$ and $\gamma$ are fixed, but it is not unique. Alternatively, equally reasonable identifications of $k(r)$ would lead to different functions $G(r)$ and thus to different RG-improved black-hole geometries. Combining these expressions gives the scale-dependent Newton constant,
\be
\label{Geffr}
G(r) = \frac{G_N r^3}{r^3 + \om G_N \hbar \left(r + \gamma G_N M \right)},
\ee
which reduces to the standard Newtonian coupling in the classical limit $\hbar \to 0$. 
Clearly, the presence of $\hbar$ signals the quantum character of the correction that the SDG/ASG approach gives to the standard general relativity theory.
In fully geometrized Planck units ($c = k_B = G_N = \hbar = 1$), this simplifies to
\be
\label{run_c}
G(r)=\frac{r^3}{r^3+\alpha(r+\gamma M)}\,.
\ee
Since the Planck length $\ell_p=\sqrt{G_N\hbar/c^3}$ and the Planck mass $m_p=\sqrt{\hbar c/4G_N}$, then in geometrized units we have $\ell_p=1$, and $m_p=1/2$, hence it follows that $r$ is measured in Planck lengths, and $M$ is the dimensionless geometric mass linked to the physical mass via $M_{\rm phys}=2M m_p$.

In the ASG literature, the parameter $\om$ is always taken to be positive. In fact, this ensures that $dG/dk<0$, $G(k) \geq 0$, and $G(k) \to 0$ as $k \to \infty$, preserving asymptotic safety and avoiding UV divergences. Negative $\om$ values, in contrast, lead to a divergence of $G(k)$ near the Planck scale, sign changes for $k > k_{\mathrm{Planck}}$, and $G(k) \to 0^{-}$ at high energies. In particular, there is no regime in which a positive Newton coupling approaches zero in the ultraviolet, so the RG flow does not realise the non-Gaussian fixed point of asymptotically safe gravity. As a result, the UV behaviour of $G(k)$ lies outside asymptotically safe gravity in the strict sense. Nevertheless, a theory with $\alpha<0$ can still be classified as an SDG theory, and this is precisely what happens for the Planck-star metric considered here (see \cite{Scardigli2023PRD}). When the free parameters of an SDG/ASG-inspired metric~\cite{Bonanno2000PRD, Bonanno2004, Liu2018, Platania2019, Koch:2014cqa, Bonanno2016} are fixed by matching its effective Newtonian potential to the one derived in general relativity treated as an effective field theory (see, e.g., \cite{Duff1974, Donoghue1994, Hamber1995PLB, Khriplovich2002JETP, Bjerrum2003PRD, Khriplovich2004JETP, Bjerrum2003PRDa, Bjerrum2015PRL, Akhundov2008EJTP, Kiefer, Donoghue2019, Shapiro2022, Frob2022}), a negative $\alpha$ emerges naturally \cite{Scardigli2023PRD}. This procedure places the resulting geometry outside the standard ASG class, but firmly within SDG, and throughout this work, we explicitly work in this $\alpha<0$ SDG regime. It is worth noticing that in this setting, the choice $\om < 0$ leads directly, without additional assumptions, to a specific black hole metric reproducing key features of the so-called Planck stars. While the Planck-star concept was initially proposed on heuristic grounds \cite{RovelliPS}, it arises here as a natural consequence of the SDG framework, as we shall demonstrate in the next Section.

\section{Prolegomena on Planck stars}

According to the discussion in the previous section, the RG-improved Schwarzschild line element in Boyer–Lindquist coordinates reads (see \cite{Bonanno2000PRD, Lambiase2022PRD, Scardigli2023PRD})
\begin{equation}\label{LE}
ds^2=-F(r)dt^2+\frac{dr^2}{F(r)}+r^2 d\vartheta^2+r^2\sin^2{\vartheta}d\varphi^2,
\qquad\quad
F(r)=1-\frac{2MG(r)}{r}=1-\frac{2 M r^2}{r^3 +  \om (r + \gamma  M)}.
\end{equation}  
The usual Schwarzschild metric is recovered for $\alpha = 0$. Two important limiting cases are worth noting. At low energy scales, i.e. as $r\to\infty$ or $k\to 0$, we recover again the standard Schwarzschild metric 
\begin{equation}
F(r \to \infty) \simeq 1 - \frac{2M}{r} ,
\end{equation}
and the behavior of $F$ is independent of the values of $\om$ and $\gamma$. At high energy scales, i.e. $r\to 0$ or $k\to\infty$, the metric \eqref{LE} goes over to   
\begin{equation}
F(r \to 0) \simeq 1 - \frac{2 r^2}{\om \gamma},
\end{equation}
which, depending on the sign of $\om \gamma$, yields a regular de Sitter ($\om\gamma > 0$) or anti–de Sitter ($\om\gamma < 0$) core \cite{Scardigli2023PRD}. In other words, the RG-improved metric interpolates between Schwarzschild at large $r$ and a (A)dS-like core at small $r$. When $\gamma = 0$, the lapse function becomes
\begin{equation}
F(r)=1-\frac{2M r}{r^2+\alpha}
\label{7}
\end{equation}
with the small-$r$ limit
\begin{equation}
  F(r \to 0) \simeq 1 - \frac{2Mr}{\om}.
\label{8}
\end{equation}
Although $F(0) = 1$ is finite, the Kretschmann scalar
\begin{equation}
K^{\mu\nu\rho\lambda}K_{\mu\nu\rho\lambda}=\frac{16 M^{6}\left(3r^{8}-2 \alpha r^{6}+13\alpha^{2}r^{4}+4\alpha^{3}r^{2}+2 \alpha^{4}\right)}{r^2\left(r^{2}+\alpha\right)^{6}}
\end{equation}
is singular at $r = 0$, indicating a conic singularity. To fix $\om$, we compare the Newtonian potential from the SDG metric (for the sake of clarity, we retain $G_N$ and $\hbar$ explicitly in the next two equations)
\be
\label{Newrun}
V^{SDG}(r) = - \frac{G_N M m}{r}\left[1 - \frac{\om G_N \hbar}{r^2} - \frac{\gamma \om G_N^2 \hbar M}{r^3} + {\cal O}\left(\frac{G_N^2 \hbar^2}{r^4}\right)\right]\,
\ee
with the quantum-corrected Newtonian potential from GR as an effective field theory
\be
\label{Newquan}
  V^{QGR}(r) \ = \
	- \frac{G_N M m}{r}\left[1 \ + \ \frac{41}{10\pi}\frac{G_N \hbar}{r^2} \ + \ \dots\right]\,.
\ee 
Matching coefficients yields \cite{Lambiase2022PRD,Scardigli2023PRD}
\be
\om = - \frac{41}{10\pi}\,.
\label{om}
\ee 
To gain insights into the historical evolution of the numerical value of $\alpha$, we refer the reader to \cite{Scardigli2023PRD, Hamber1995PLB, Bjerrum2003PRD, Khriplovich2002JETP, Bjerrum2003PRDa, Khriplovich2004JETP, Akhundov2008EJTP, Kiefer, Bjerrum2015PRL, DonoghueJPG2015, Batic2016EPJC, Batic2017EPL, Scardigli2017PLB}. The parameter $\gamma$ can be determined following the classical general relativistic arguments first presented in \cite{Bonanno2000PRD} and subsequently reinforced by other works (e.g., \cite{Koch:2014cqa}). These considerations lead to the specific choice $\gamma = 9/2$, which we also adopt here, in agreement with previous studies \cite{Bonanno2000PRD, Scardigli2023PRD, Lambiase2023EPJC}. Throughout this work, we restrict to $\gamma > 0$, for which the qualitative features of the solutions remain essentially unchanged across different positive values. This choice is physically motivated, as a positive $\gamma$ ensures a regular high-energy core and avoids pathologies such as reversed gravitational coupling near the origin. In a few specific instances, however, we also comment on the special case $\gamma = 0$, as previously discussed.

\begin{figure}[h]
\centering
\includegraphics[scale=1.2]{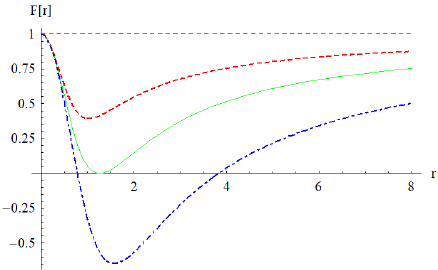}
\caption{Lapse function $F(r)$ for $\om > 0$: Horizons correspond to the zeros of $F(r)$. The curves are shown for a fixed $\om > 0$ and increasing values of the mass parameter $M$: $0 < M_{\text{red}} < M_{\text{green}} \equiv M_c < M_{\text{blue}}$. At the critical mass $M = M_c$ (green curve), the two horizons merge, corresponding to the extremal black hole configuration. For $M > M_c$ (blue curve), the lapse exhibits two distinct horizons $r_-$ and $r_+$, where $F(r_-) = F(r_+) = 0$.}
\label{fig1}
\end{figure}

Regarding the horizon and singularity structure, for any $\om > 0$, Fig.~\ref{fig1} shows that the lapse function \eqref{LE} remains regular in the physical domain $r \geq 0$. Depending on the mass $M$ of the central object, the metric may exhibit two distinct horizons ($r_-$, $r_+$), a pair of coincident horizons, or no horizons at all. A detailed mathematical proof of these general properties for $\alpha > 0$, together with explicit expressions for $r_\pm$, can be found in \cite{Hassannejad2025PRD} (Appendix E). In the simpler special case $\gamma = 0$, the two horizons, i.e. the outer event horizon $r_+$ and the inner Cauchy horizon $r_-$, are given analytically by $r_\pm = M \pm \sqrt{M^2 - \alpha}$, satisfying $0 < r_- < r_+$ and $0 < \alpha \leq M^2$ as before.

\begin{figure}[b]
	\includegraphics[scale=.4]{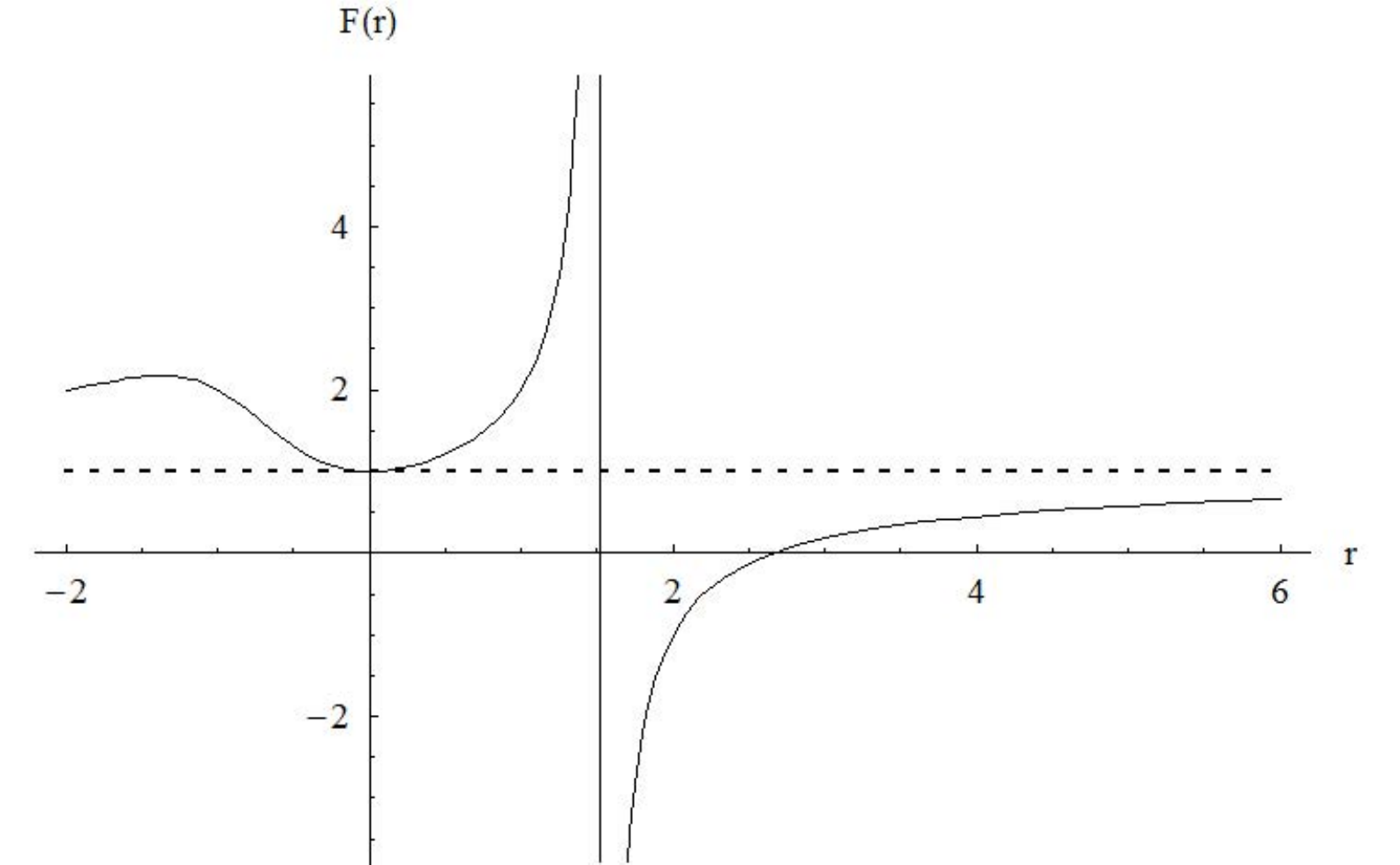}
	\centering
	\protect\caption{Lapse function $F(r)$ for $\alpha < 0$. In the physical region $r > 0$, the spacetime always contains a curvature singularity located at $r = r_0 > 0$ and a single event horizon at $r = r_h > r_0$, where $F(r_h) = 0$. The horizon thus always encloses the singularity, ensuring the absence of naked singularities.}
	\label{fig2}
\end{figure}
\begin{figure}[b]
	\includegraphics[scale=.4]{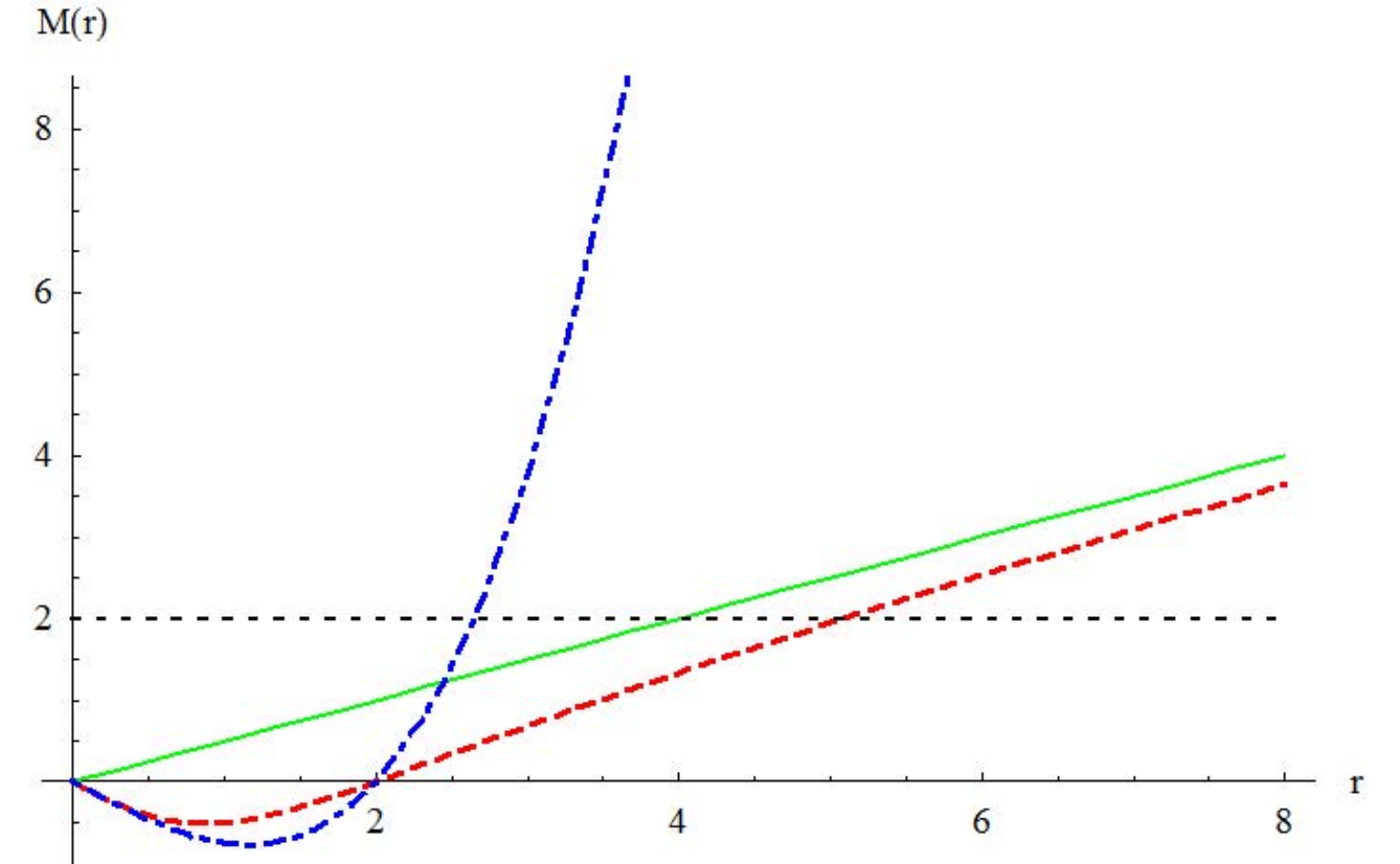}
	\centering
	\protect\caption{Horizon mass function $M(r_h)$ (red dashed line), singularity mass function $M(r_0)$ (blue dot–dashed line), and Schwarzschild mass function $M = r_{\mathrm{SCH}}/2$ (green solid line). The horizontal black dashed line represents an arbitrary positive mass $M > 0$. For any such $M$, the intersection with the blue curve occurs at $r_0(M)$ before that with the red curve at $r_h(M)$, ensuring $r_0(M) < r_h(M)$ and thus the absence of naked singularities.}
	\label{fig3}
\end{figure}

For $\alpha<0$, the analysis in \cite{Scardigli2023PRD, Hassannejad2025PRD}, which is summarised in Fig.~\ref{fig2}, shows that the lapse function $F(r)$ always possesses a single curvature singularity at a positive radial coordinate $r_0>0$ and exactly one event horizon $r_h$, with the ordering $0<r_0<r_h$. The geometry is therefore non-regular because curvature invariants diverge at the finite radius $r_0$, so the classical Schwarzschild singularity is not removed but displaced from $r = 0$ to a spherical surface of finite areal radius hidden inside the horizon. Furthermore, the singularity radius $r_0$ is the positive root of
\begin{equation}\label{SE}
  r_0^3 + \om \big(r_0 + \gamma M\big) = 0,
\end{equation}
and corresponds to a true curvature singularity, as confirmed by the divergence of the Kretschmann invariant:
\begin{align}
\label{djjlwqqw}
&R_{\mu\nu\alpha\beta}R^{\mu\nu\alpha\beta}=\notag\\
&\frac{256 M^6 \left[ \left(r^3+\om(r +\gamma M) \right)^4
+\left(r^3+\om(r +\gamma M)\right)^2 \left(r^3-\om(r+2\gamma M)\right)^2+
\left(r^6-\om r^3(3r+7\gamma M)+(\om \gamma M)^2\right)^2 \right]}{\big[r^3+\om(r +\gamma M)\big]^6}\,.
\end{align}
Indeed, if $r_0$ satisfies $r^3 + \om r + \om\gamma M = 0$, the denominator in \eqref{djjlwqqw} vanishes, and the curvature invariant diverges as $r \to r_0$. Following \cite{Hassannejad2025PRD} (see in particular Appendix E), the single event horizon $r_h$, which always lies outside the singularity at $r_0$, is given by the positive root of
\begin{equation}\label{HE}
  r_h^3 - 2Mr_h^2 + \alpha r_h +\alpha\gamma M=0\,. 
\end{equation}
The strict ordering $r_0 < r_h$ guarantees that the singularity is always hidden within the event horizon, ensuring that the spacetime contains no naked singularities for $\alpha < 0$. For $\om < 0$, it is instructive to examine the large-$M$ asymptotics of both the event horizon radius $r_h(M)$ and the singularity radius $r_0(M)$. Starting with the event horizon, $r_h(M)$ is given by the solution of \eqref{HE}. In the limit $M \gg 1$, we find
\be
  r_h(M) \ \simeq \ 2M \ + \ \frac{(2+\gamma)|\om|}{4M}\,,
\ee 
so that the standard Schwarzschild relation $r_h \approx 2M$ is recovered at leading order. For the singularity radius, obtained from \eqref{SE}, the large-$M$ behavior is
\be
\label{r0}
r_0(M) \ \simeq \ (\gamma |\om| M)^{1/3} \ + \ \frac{|\om|}{3(\gamma |\om| M)^{1/3}}\,.
\ee
The relationship between these two roots becomes even clearer by inspecting Fig.~\ref{fig3}, which plots the corresponding mass functions
\be
M(r_h) = \left. \frac{r^3 - |\om| r}{2r^2 + |\om|\gamma} \right|_{r=r_h}\,, \quad\quad\quad
M(r_0) = \left. \frac{r^3 - |\om| r}{|\om|\gamma} \right|_{r=r_0}\,,
\label{hmf}
\ee  
together with the standard Schwarzschild relation $M=r_{SCH}/2$. In the diagram, any horizontal line (black dashed) corresponding to a fixed $M > 0$ intersects the blue curve (singularity) before the red curve (horizon), confirming that $r_0(M) < r_h(M)$ for all $M > 0$. Consequently, the curvature singularity is always hidden inside the horizon, ensuring the absence of naked singularities. Both mass functions vanish at the same critical radius $r_c = \sqrt{|\om|}$ with $r_c^2 = |\om|$. As shown in Table~\ref{table:event}, increasing the mass $M$ causes $r_h$ to asymptotically approach the Schwarzschild radius, indicating convergence toward the classical black hole limit.
\begin{table}[ht]
\centering
\caption{Representative numerical values of the curvature singularity position $r_0$ and event horizon radius $r_h$ for various values of the mass parameter $M$ and $\gamma=9/2$.}
\label{table:event}
\vspace*{1em}
\begin{tabular}{||c|c|c|c||}
\hline\hline
$M$ &              $r_0$             & $r_h$           \\ [0.5ex]
\hline\hline
$1.0$              & $2.044060591$   & $3.055929964$   \\
$2.0$              & $2.464097888$   & $4.785579214$   \\
$5.0$              & $3.226034006$   & $10.39715780$   \\
$10$               & $3.998895843$   & $20.20838866$   \\
$10^3$             & $18.06600745$   & $2000.002121$   \\[1ex]
\hline\hline 
\end{tabular}
\end{table}

It is also useful to quantify how the SDG corrections to the geometry decouple in the large-mass limit in the strong-field region relevant for QNMs. Introducing the dimensionless variable $x = r/M$ and rewriting the lapse function \eqref{LE} in terms of $x$ yields
\begin{equation}
F(r) = 1 - \frac{2}{x}\frac{1}{1 + \dfrac{\alpha(\gamma + x)}{M^{2} x^{3}}}, \qquad r = x M.
\end{equation}
For $M \gg 1$ and fixed $x = \mathcal{O}(1)$ one then finds the large-mass expansion
\begin{equation}\label{expand}
F(r) = 1 - \frac{2}{x} + \frac{2\alpha(\gamma + x)}{x^{4} M^{2}} + \mathcal{O}\left(\frac{1}{M^4}\right).
\end{equation}
Thus, in the photon sphere region $r \sim \mathcal{O}(M)$, the deviation from the Schwarzschild lapse is suppressed as $|\delta F| \propto |\alpha|/M^{2}$ in Planck units.

To interpret the metric \eqref{LE} as that of a Planck star, it is crucial that the behaviour of the quantity $r_0(M)$ given in \eqref{r0}, which determines the size of the black hole’s central core. For masses $M$ much larger than the Planck mass ($M \gg m_p$), the central curvature singularity is located at $r = r_0$ and possesses a finite, nonzero size $r_0 \simeq (\gamma |\om| G^2 \hbar M)^{1/3} > 0$ (we restore $G$ and $\hbar$ here for clarity). This is in stark contrast with the standard Schwarzschild black hole, where the singularity is point-like. The nonzero extent of $r_0$ arises entirely from quantum effects, as it vanishes in the classical limit $\hbar \to 0$. Most importantly, if we assume that the entire collapsing mass $M$ is concentrated within a compact central sphere of radius $r_0$, we can compute its (non-covariant) volume and thus estimate the matter density as perceived by a distant observer. The core radius is given by
\be
r_0 = (\gamma |\om| \ell_p^2 G M)^{1/3} \ = 
\ (\gamma|\om|/2)^{\frac{1}{3}} \left(\frac{M}{m_p}\right)^{\frac{1}{3}}\ell_p\,, \nonumber
\ee 
which immediately yields the density
\be
\varrho = \frac{M}{V_{core}} = \frac{3}{2\pi}\frac{m_p}{\gamma |\om| \ell_p^3} \simeq 
\frac{m_p}{2\gamma|\om| \ell_p^3} = \frac{\varrho_{Planck}}{2\gamma|\om|},
\label{size}
\ee
where we used the definitions $G\hbar = \ell_p^2$, $2G m_p = \ell_p$, and $\varrho_{\mathrm{Planck}} = m_p / \ell_p^3$. This result shows that, from the point of view of a distant observer, the central core of our black hole has both a finite size and a finite effective density, the latter being of order the Planck density. These properties match the heuristic picture of the so-called Planck stars, first introduced in \cite{RovelliPS} (see also \cite{BonannoCasadio}) on the basis of general physical arguments, and we shall therefore refer to the configuration \eqref{LE} as a Planck-star geometry. At the same time, it is important to stress that, unlike some regular Planck-star models discussed in the literature, the spacetime studied here is not free of singularities. As shown above, the Kretschmann scalar diverges at $r=r_0>0$, so the singularity is extended but not resolved, and our static analysis does not implement any dynamical quantum bounce. Remarkably, many of the qualitative features described in \cite{RovelliPS, BonannoCasadio} emerge naturally from the metric \eqref{LE} without additional assumptions. The finite extent of the core is purely quantum in origin, i.e. it vanishes in the classical limit, and is plausibly the result of the action of the Heisenberg uncertainty principle, which forbids matter from being compressed into a point of zero size. Furthermore, this central quantum kernel could, in principle, serve as a repository for the information absorbed by the black hole, offering a potential pathway to resolve the information paradox. All these considerations apply strictly for $\gamma > 0$. Last but not least, from the point of view of classical GR, the RG-improved metric \eqref{LE} can always be rewritten as a solution of Einstein's equations with an effective source, $G_{\mu\nu} = 8\pi G_N T_{\mu\nu}^{(eff)}$, where $T_{\mu\nu}^{(eff)}$ encodes the back-reaction of the running coupling $G(r)$. This tensor should be understood as an effective stress–energy tensor summarising short-distance quantum-gravity corrections, rather than as the energy–momentum tensor of a standard classical fluid. In particular, near the Planck-density core, $T_{\mu\nu}^{(eff)}$ does not satisfy the usual classical energy conditions, a behaviour that is common in quantum-gravity–inspired black-hole geometries. In the present work, we remain agnostic about the microscopic degrees of freedom underlying $T_{\mu\nu}^{(eff)}$ and focus instead on the geometric structure of the Planck-star spacetime and its imprint on the QNM spectrum.

\section{Quasinormal modes of Planck stars}

In the presence of a spherically symmetric metric such as \eqref{LE}, the equation governing a massless Klein-Gordon field (assumed to have a time dependence of the form $e^{-i\omega t}$ and an angular component described by spherical harmonics) is as follows \cite{Batic2019EPJC}
\begin{equation}\label{ODE01}
    F(r)\frac{d}{dr}\left(F(r)\frac{d\psi_{\omega\ell\epsilon}}{dr}\right)+\left[\omega^2-U_\epsilon(r)\right]\psi_{\omega\ell\epsilon}(r)=0,\qquad
    U_\epsilon(r)=F(r)\left[\frac{\epsilon}{r}\frac{dF}{dr}+\frac{\ell(\ell+1)}{r^2}\right],\qquad
    \epsilon=1-s^2
\end{equation}
with $\ell = 0, 1, 2, \ldots$ and $\epsilon = -1$ (massless scalar perturbation $s=0$), $\epsilon = 0$ (electromagnetic perturbation $s=1$), and $\epsilon = 3$ (vector-type gravitational perturbation $s=2$). Notice that in the present case $F(r)$ is given by the second equation in \eqref{LE}. By means of the substitution $x=r/r_h$, the above equation can be recast in the equivalent form
\begin{equation}\label{ourODE}
F(x)\frac{d}{dx}\left(F(x)\frac{d\psi_{\omega\ell\epsilon}}{dx}\right)+\left[r_h^2\omega^2-V_\epsilon(x)\right]\psi_{\omega\ell\epsilon}(x)=0,\qquad
F(x)=1-\frac{2Mr_h^2 x^2}{r_h^3 x^3 +\alpha(r_h x+\gamma M)}
\end{equation} 
with an effective potential given by
\begin{equation}
V_\epsilon(x)=F(x)\left[\frac{\epsilon}{x}\frac{dF}{dx}+\frac{\ell(\ell+1)}{x^2}\right].
\end{equation}
In the following analysis, we compute the QNMs of the spectral problem in \eqref{ourODE}. For this purpose, we represent $\omega$ as $\omega = \omega_R + i\omega_I$, where $\omega_I < 0$ ensures that the perturbation is dumped in time. The boundary conditions are set so that the radial field exhibits inward radiation at the event horizon and outward radiation at spatial infinity. This necessitates a thorough examination of the solution asymptotic behaviour in \eqref{ourODE}, both near the event horizon ($x \to 1^{+}$) and at large spatial distances ($x \to +\infty$). Moreover, to compute the QNMs using the SM, we must recast the differential equation \eqref{ourODE} and the appropriate boundary conditions over the compact interval $[-1,1]$. This adjustment is necessary because the method expands the regular part of the eigenfunctions in terms of Chebyshev polynomials. We begin by observing that the event horizon corresponds to a simple zero of the function $F(x)$ defined in \eqref{ourODE}. Furthermore, imposing the condition $F(1) = 0$, which is equivalent to $\alpha M\gamma=r_h[r_h(2M-r_h)-\alpha]$, enables us to recast $F(x)$ in the form
\begin{equation}
F(x)=1-\frac{2Mr_h^2 x^2}{r_h^3 x^3+\alpha r_h(x-1)+r_h^2(2M-r_h)}.
\end{equation}
To establish the QNM boundary conditions at the event horizon and at infinity, we first need to determine the asymptotic behaviour of the radial solution $\psi_{\omega\ell\epsilon}$ as $x \to 1^{+}$ and as $x \to +\infty$. We can then extract the QNM boundary conditions from this asymptotic data. Concerning the asymptotic behaviour as $x\to 1^+$, it is convenient to reformulate \eqref{ourODE} in the form
\begin{equation}\label{ODEZ}
\frac{d^2\psi_{\omega\ell\epsilon}}{dx^2}+p(x)\frac{d\psi_{\omega\ell\epsilon}}{dx}+q(x)\psi_{\omega\ell\epsilon}(x)=0,\quad
p(x)=\frac{F^{'}(x)}{F(x)},\quad
q(x)=\frac{x_h^2\Omega^2-V_\epsilon(x)}{F^2(x)}.   
\end{equation}
Since $p$ and $q$ have poles of order one and two at $x = 1$, respectively, this point is classified as a regular singular point of \eqref{ODEZ}, according to Frobenius theory \cite{Ince1956}. Hence, we can construct solutions of the form
\begin{equation}
\psi_{\omega\ell\epsilon}(x)=(x-1)^\rho\sum_{\kappa=0}^\infty a_\kappa(x-1)^\kappa.
\end{equation}
The leading behavior at $x=1$ is represented by the term $(x-1)^\rho$ where $\rho$ is determined by the indicial equation
\begin{equation}\label{indicial}
\rho(\rho-1)+P_0\rho+Q_0=0
\end{equation}
with
\begin{equation}
P_0=\lim_{x\to 1}(x-1)p(x)=1,\qquad
Q_0=\lim_{x\to 1}(x-1)^2 q(x)=\left(\frac{r_h \omega}{F^{'}(1)}\right)^2.
\end{equation}
The roots of \eqref{indicial} are $\rho_\pm = \pm i r_h\omega/F^{'}(1)$ and the correct QNM boundary condition at $x=1$ reads
\begin{equation}\label{QNMBCz1}
\psi_{\omega\ell\epsilon}\underset{{x\to 1^+}}{\longrightarrow} (x-1)^{-i r_h a\omega},\quad a=\frac{1}{F^{'}(1)}=\frac{2Mr_h}{3r_h^2-4Mr_h+\alpha}.
\end{equation}
From the above expression, we immediately see that $a$ is always positive since $\alpha<0$. Regarding the asymptotic behaviour as $x\to+\infty$, we start by observing that
\begin{equation}
p(x) = \sum_{\kappa=0}^\infty\frac{\mathfrak{f}_\kappa}{x^k} = \mathcal{O}\left(\frac{1}{x^2}\right), \qquad
q(x) = \sum_{\kappa=0}^\infty\frac{\mathfrak{g}_\kappa}{z^k}=r_h^2\omega^2+\frac{4Mr_h\omega^2}{x}+\mathcal{O}\left(\frac{1}{x^2}\right).
\end{equation}
Consequently, the asymptotic behaviour of the solutions to equation \eqref{ODEZ}) can be deduced using the method outlined in \cite{Olver1994MAA}. Given that at least one of the coefficients $\mathfrak{f}_0$, $\mathfrak{g}_0$, $\mathfrak{g}_1$ is nonzero, a formal solution to \eqref{ODEZ} is represented by \cite{Olver1994MAA}
\begin{equation}\label{olvers}
\psi^{(j)}_{\omega\ell\epsilon}(x) = x^{\mu_j}e^{\lambda_j x}\sum_{\kappa=0}^\infty\frac{a_{\kappa,j}}{x^\kappa}, \qquad j \in \{1,2\},
\end{equation}
where $\lambda_1$, $\lambda_2$, $\mu_1$ and $\mu_2$ are the roots of the characteristic equations
\begin{equation}\label{chareqns}
   \lambda^2+\mathfrak{f}_0\lambda+\mathfrak{g}_0=0,\quad
   \mu_j=-\frac{\mathfrak{f}_1\lambda_j+\mathfrak{g}_1}{\mathfrak{f}_0+2\lambda_j}.
\end{equation}
A straightforward computation shows that $\lambda_\pm = \pm ir_h\omega$ and $\mu_\pm = \pm 2iM\omega$. As a result, the QNM boundary condition at space-like infinity can be expressed as
\begin{equation}\label{QNMBCzinf}
    \psi_{\omega\ell\epsilon}\underset{{x\to +\infty}}{\longrightarrow} x^{2iM\omega}e^{i r_h\omega x}.
\end{equation}
At this point, we transform the radial function $\psi_{\omega\ell\epsilon}(x)$ into a new radial function $\Phi_{\omega\ell\epsilon}(x)$ such that the QNM boundary conditions are automatically implemented and  $\Phi_{\omega\ell\epsilon}(x)$ is regular at $x = 1$ and at space-like infinity. To this aim, we consider the transformation
\begin{equation}\label{Ansatz}
\psi_{\omega\ell\epsilon}(x) = x^{i(2M+ar_h)\omega}(x-1)^{-iar_h\omega}e^{ir_h\omega(x-1)} \Phi_{\omega\ell\epsilon}(x).
\end{equation}
If we substitute \eqref{Ansatz} into \eqref{ourODE}, we end up with the following ordinary differential equation for the radial eigenfunctions, namely
\begin{equation}\label{ODEznone}
    P_2(x)\Phi^{''}_{\omega\ell\epsilon}(x) + P_1(x)\Phi^{'}_{\omega\ell\epsilon}(x) + P_0(x)\Phi_{\omega\ell\epsilon}(x) = 0
\end{equation}
with
\begin{eqnarray}
P_2(x)&=&x^2(x-1)^2 F^2(x),\\
P_1(x)&=&x(x-1)F(x)\left\{x(x-1)F^{'}(x)+i\omega F(x)\left[2r_h x^2+2x(2M-r_h)-2ar_h-4M\right]\right\},\\
P_0(x)&=&-\omega^2 Q_+(x)Q_{-}(x)+i\omega F(x)L(x)-x^2(x-1)^2 V_\epsilon(x),\\
Q_\pm(x)&=&F(x)[(x-1)(r_h x+2M)-ar_h]\pm r_h x(x-1),\\
L(x)&=&x(x-1)[r_h x^2+(2M-r_h)x-2M-ar_h]F^{'}(x)-F(x)[2Mx^2-(2x-1)(2M+ar_h)].
\end{eqnarray}
Let us now introduce the transformation $x=2/(1-y)$ mapping the point at infinity and the event horizon to $y = 1$ and $y = -1$, respectively. Furthermore, a dot denotes differentiation with respect to the new variable $y$. Then, equation \eqref{ODEznone} becomes
\begin{equation}\label{ODEynone}
    S_2(y)\ddot{\Phi}_{\omega\ell\epsilon}(y) + S_1(y)\dot{\Phi}_{\omega\ell\epsilon}(y) + S_0(y)\Phi_{\omega\ell\epsilon}(y) = 0,
\end{equation}
where
\begin{eqnarray}
  S_2(y) &=&(1+y)^2 F^2(y), \label{S2onone} \\
  S_1(y) &=& i\omega\frac{1+y}{(1-y)^2}F^2(y)\left[8r_h+4(2M-r_h)(1-y)-2(2M+ar_h)(1-y)^2\right]\nonumber\\
  &&-2\frac{(1+y)^2}{1-y}F^2(y)+(1+y)^2 F(y)\dot{F}(y), \label{S1onone}\\
  S_0(y) &=& \omega^2\Sigma_2(y)+i\omega\Sigma_1(y)+\Sigma_0(y) \label{S0onone}
\end{eqnarray}
with
\begin{eqnarray}
\Sigma_2(y) &=& \frac{4r_h^2(1+y)^2}{(1-y)^4}-\frac{F^2(y)}{(1-y)^4}\left\{2M(1-y^2)-r_h[ay^2-2y(1+a)+a-2]\right\}^2,\\
\Sigma_1(y) &=&F(y)\left\{
\frac{1+y}{(1-y)^2}[4r_h+2(2M-r_h)(1-y)-(2M+ar_h)(1-y)^2]\dot{F}(y)\right.\nonumber\\
&&-\left.\frac{F(y)}{(1-y)^2}[8M-(2M+ar_h)(3+y)(1-y)]
\right\},\\
\Sigma_0(y)&=&-\frac{4(1+y)^2}{(1-y)^4}V_\epsilon(y).
\end{eqnarray}
Notice that we must also require that $\Phi_{\omega\ell\epsilon}(y)$ is regular at $y=\pm 1$. As a result of the transformation introduced above, we have 
\begin{equation}\label{fv}
F(y)=1-\frac{8Mr_h^2(1-y)}{8r_h^3+2\alpha r_h(1-y)^2-r_h(r_h^2-2Mr_h+\alpha)(1-y)^3}, \quad 
  V_\epsilon(y) = \frac{(1-y)^2}{4}F(y)\left[\epsilon (1-y)\dot{F}(y)+\ell(\ell+1)\right].
\end{equation}
\begin{table}
\caption{Classification of the points $y=\pm 1$ for the relevant functions defined by   (\ref{S2onone})--(\ref{S0onone}), and (\ref{fv}),. The abbreviations $z$ ord $n$ and $p$ ord $m$ stand for zero of order $n$ and pole of order $m$, respectively.}
\begin{center}
\begin{tabular}{ | c | c | c | c | c | c | c | c }
\hline
$y$  & $F(y)$  & $V_\epsilon(y)$ & $S_2(y)$ & $S_{1}(y)$ & $S_{0}(y)$\\ \hline
$-1$ & z \mbox{ord} 1 & z \mbox{ord} 1 & z \mbox{ord} 4& z \mbox{ord} 3 & z \mbox{ord} 3 \\ \hline
$+1$ & $+1$  & z \mbox{ord} 2 & $4$ & p \mbox{ord} 2 & p \mbox{ord} 2\\ \hline
\end{tabular}
\label{tableEinsnone}
\end{center}
\end{table}
Table~\ref{tableEinsnone} indicates that the coefficients of the differential equation \eqref{ODEynone} share a common zero of order $2$ at $y = -1$ while $y = 1$ is a pole of order $4$ for the coefficient $S_0$. Hence, in order to apply the SM, we need to multiply \eqref{ODEynone} by $(1-y)^4/(1+y)^2$. As a result, we end up with the following differential equation
\begin{equation}\label{ODEhynone}
M_2(y)\ddot{\Phi}_{\omega\ell\epsilon}(y) + M_1(y)\dot{\Phi}_{\omega\ell\epsilon}(y) + M_0(y)\Phi_{\omega\ell\epsilon}(y) = 0,
\end{equation}
where
\begin{equation}\label{S210honone}
M_2(y) = \frac{(1-y)^2}{1+y} F^2(y), \qquad
M_1(y) = i\omega N_1(y)+N_0(y), \qquad
M_0(y) = \omega^2 C_2(y)+i\omega C_1(y)+C_0(y)
\end{equation}
with
\begin{eqnarray}
N_1(y) &=&\frac{2F^2(y)}{(1+y)^2}\left[4r_h+2(2M-r_h)(1-y)-(ar_h+2M)(1-y)^2\right],\label{N1}\\
N_0(y) &=&\frac{1-y}{1+y}\left[(1-y)\dot{F}(y)-2F(y)\right],\label{N0}\\
C_2(y) &=& \frac{4r_h^2}{(1+y)(1-y)^2}-\frac{F^2(y)}{(1+y)^3(1-y)^2}\left\{2M(1-y^2)-r_h[ay^2-2y(1+a)+a-2]\right\}^2,\label{C2}\\
C_1(y) &=&\frac{F(y)}{(1+y)^3}\left\{(1+y)\dot{F}(y)\left[4r_h+2(2M-r_h)(1-y)-(ar_h+2M)(1-y)^2\right]\right.\nonumber\\
&&\left.-F(y)\left[8M-(3+y)(1-y)(ar_h+2M)\right]\right\},\label{C1}\\
C_0(y) &=& -\frac{4V_\epsilon(y)}{(1+y)(1-y)^2}.\label{C0}
\end{eqnarray}
\begin{eqnarray}
    &&\lim_{y\to 1^{-}}M_2(y)=0=\lim_{y\to -1^{+}}M_2(y),\\
    &&\lim_{y\to 1^{-}}M_1(y)=2ir_h\omega,\quad
    \lim_{y\to -1^{+}}M_1(y)=i\omega\Lambda_1+\Lambda_0,\\
    &&\lim_{y\to 1^{-}}M_0(y)=A_2\omega^2 +A_0,\quad
     \lim_{y\to -1^{+}}M_0(y)=B_2\omega^2+i\omega B_1+B_0,
\end{eqnarray}
where
\begin{eqnarray}
\Lambda_1 &=&\frac{4Mr_h-3r_h^2-\alpha}{M},\quad
\Lambda_0 = \left(\frac{\Lambda_1}{2r_h}\right)^2,\quad
A_2=\frac{2M(r_h^3+6Mr_h^2-8M^2 r_h+2M\alpha)}{3r_h^2-4Mr_h+\alpha},\quad
A_0=-\frac{\ell(\ell+1)}{2},\label{coef1}\\
B_2&=&\frac{9r_h^5-9Mr_h^4-6(3M^2-\alpha)r_h^3+M(16M^2+\alpha)r_h^2-\alpha(8M^2-\alpha)r_h+M\alpha^2}{M(3r_h^2-4Mr_h+\alpha)},\label{coef2}\\
B_1&=&\frac{3r_h^2-4Mr_h+\alpha}{4Mr_h^2}B_2,\quad
B_0=-\frac{3r_h^2-4Mr_h+\alpha}{8M^2 r_h^2}\left\{3\epsilon r_h^2+2M[\ell(\ell+1)-2\epsilon]+\alpha\epsilon\right\}.\label{coef3}
\end{eqnarray}
In the final step leading to the application of the SM, we recast the differential equation \eqref{ODEhynone} into the following form
\begin{equation}\label{TSCH}
  L_0\left[\Phi_{\omega\ell\epsilon}, \dot{\Phi}_{\omega\ell\epsilon}, \ddot{\Phi}_{\omega\ell\epsilon}\right] +  i L_1\left[\Phi_{\omega\ell\epsilon}, \dot{\Phi}_{\omega\ell\epsilon}, \ddot{\Phi}_{\omega\ell\epsilon}\right]\omega +  L_2\left[\Phi_{\omega\ell\epsilon}, \dot{\Phi}_{\omega\ell\epsilon}, \ddot{\Phi}_{\omega\ell\epsilon}\right]\omega^2 = 0
\end{equation}
with
\begin{eqnarray}
L_0\left[\Phi_{\omega\ell\epsilon}, \dot{\Phi}_{\omega\ell\epsilon}, \ddot{\Phi}_{\omega\ell\epsilon}\right] &=& L_{00}(y)\Phi_{\omega\ell\epsilon} + L_{01}(y)\dot{\Phi}_{\omega\ell\epsilon} + L_{02}(y)\ddot{\Phi}_{\omega\ell\epsilon},\label{L0none}\\
L_1\left[\Phi_{\omega\ell\epsilon}, \dot{\Phi}_{\omega\ell\epsilon}, \ddot{\Phi}_{\omega\ell\epsilon}\right] &=&L_{10}(y)\Phi_{\omega\ell\epsilon} + L_{11}(y)\dot{\Phi}_{\omega\ell\epsilon} + L_{12}(y)\ddot{\Phi}_{\omega\ell\epsilon}, \label{L1none}\\
L_2\left[\Phi_{\omega\ell\epsilon}, \dot{\Phi}_{\omega\ell\epsilon}, \ddot{\Phi}_{\omega\ell\epsilon}\right] &=&L_{20}(y)\Phi_{\omega\ell\epsilon} + L_{21}(y)\dot{\Phi}_{\omega\ell\epsilon} + L_{22}(y)\ddot{\Phi}_{\omega\ell\epsilon}.\label{L2none}
\end{eqnarray}
Moreover, in Table~\ref{tableZweinone}, we have summarized the $L_{ij}$ appearing in (\ref{L0none})--(\ref{L2none}) and their limiting values at $y = \pm 1$.

\begin{table}
\caption{Definitions of the coefficients $L_{ij}$ and their corresponding behaviours at the endpoints of the interval $-1 \leq y \leq 1$. The symbols appearing in this table have been defined in (\ref{coef1})-(\ref{coef3}).}
\begin{center}
\begin{tabular}{ | c | c | c | c | c | c | c | c }
\hline
$(i,j)$  & $\displaystyle{\lim_{y\to -1^+}}L_{ij}$  & $L_{ij}$ & $\displaystyle{\lim_{y\to 1^-}}L_{ij}$  \\ \hline
$(0,0)$ &  $B_0$          & $C_0$                  & $A_0$\\ \hline
$(0,1)$ &  $\Lambda_0$    & $N_0$                  & $0$\\ \hline
$(0,2)$ &  $0$            & $M_2$                  & $0$\\ \hline 
$(1,0)$ &  $B_1$          & $C_1$                  & $0$\\ \hline 
$(1,1)$ &  $\Lambda_1$    & $N_1$                  & $2r_h$\\ \hline 
$(1,2)$ &  $0$            & $0$                    & $0$\\ \hline 
$(2,0)$ &  $B_2$          & $C_2$                  & $A_2$\\ \hline
$(2,1)$ &  $0$            & $0$                    & $0$\\ \hline
$(2,2)$ &  $0$            & $0$                    & $0$\\ \hline
\end{tabular}
\label{tableZweinone}
\end{center}
\end{table} 

\section{Numerical method}

In order to solve the differential eigenvalue problem \eqref{TSCH} along with the corresponding frequencies $\omega$, we have to discretize the differential operators $L_{j}[\cdot]$ with $j \in \{1,2,3\}$ defined in (\ref{L0none})-(\ref{L2none}) and in Table~\ref{tableZweinone}, respectively. Since our problem is posed on the finite interval $[-1, 1]$ without any boundary conditions, more precisely, we only require that the function $\Phi_{\omega\ell\epsilon}(y)$ be regular at $y = \pm 1$, then, it is natural to choose a Tchebyshev-type SM \cite{Trefethen2000, Boyd2000}. Namely, we are going to expand the function $y \mapsto \Phi_{\omega\ell\epsilon}(y)$ in the form of a truncated Tchebyshev series
\begin{equation}\label{eq:exp}
  \Phi_{\omega\ell\epsilon}(y)=\sum_{k=0}^{N} a_{k} T_k(y),
\end{equation}
where $N\in\ \mathbb{N}$ is kept as a numerical parameter, $\{a_{k}\}_{k=0}^{N}\ \subseteq\ \mathbb{R}$, and $\{T_k(y)\}_{k=0}^{N}$ are the Tchebyshev polynomials of the first kind
\begin{equation}
    T_k: [-1, 1]\ \longrightarrow\ [-1, 1]\,, \qquad y\ \longmapsto\ \cos\,\bigl(k\arccos y\bigr)\,.
\end{equation}
After substituting expansion \eqref{eq:exp} into the differential equation \eqref{TSCH}, we obtain an eigenvalue problem with polynomial coefficients. In order to translate it into the realm of numerical linear algebra, we employ the collocation method \cite{Boyd2000}. Specifically, rather than insisting that the polynomial function in $y$ is identically zero (a condition equivalent to having polynomial solutions for the differential problems as per equation \eqref{TSCH}), we impose a weaker requirement. This involves ensuring that the polynomial vanishes at $N+1$ strategically selected points. The number $N+1$ coincides exactly with the number of unknown coefficients $\{a_{k}\}_{k=0}^{N}$. For the collocation points, we implemented the Tchebyshev roots grid \cite{Fox1968}
\begin{equation}
  y_k= -\cos{\left(\frac{(2k+1)\pi}{2(n+1)}\right)},\quad k\in\{0, 1,\ldots,N\}.
\end{equation}
In our numerical codes, we also implemented the second option of the Tchebyshev extrema grid
\begin{equation}
  y_k=-\cos{\left(\frac{k\pi}{n}\right)},\quad k\in\{0, 1,\ldots,N\}.
\end{equation}
The users are free to choose their favourite collocation points. Notice that we used the roots grid in our computation. In any case, the theoretical performance of the two available options is known to be absolutely comparable \cite{Fox1968, Boyd2000}.

Upon implementing the collocation method, we derive a classical matrix-based quadratic eigenvalue problem, as detailed in \cite{Tisseur2001}
\begin{equation}\label{eq:eig}
  (M_{0} + iM_{1}\omega + M_{2}\omega^2)\bf{a}_\kappa =\bf{0}.
\end{equation}
In this formulation, the square real matrices $M_{j}$, each of size $(N+1)\times(N+1)$ for $j=0,1,2$, represent the spectral discretizations of the operators $L_{j}[\cdot]$, respectively. The problem \eqref{eq:eig} is solved numerically with the \textsc{polyeig} function from \textsc{Matlab}. This polynomial eigenvalue problem yields \(2(N+1)\) potential values for the spectral parameter $\omega$. To discern the physical values of $\omega$ that correspond to the black hole's QNM modes, we first overlap the root plots for various values of $N$ in equation \eqref{eq:exp}, such as $N \in \{160, 180, 200\}$. We then identify the consistent roots whose positions remain stable across these different $N$ values. To reduce rounding and other floating-point errors, we performed all computations using multiple-precision arithmetic, which is built into \textsc{Maple} and imported into \textsc{Matlab} via the \textsc{Advanpix} toolbox \cite{mct2015}. All numerical computations reported in this study have been performed with $200$ decimal digits of accuracy.

\section{Numerical results}

The SM used in this work has already been validated in a series of publications \cite{Batic2024CQG, Batic2024PRD, Batic2024EPJC, Batic2025CQG, Batic2025EPJC}. As a consistency check of the equations derived in the previous sections, we first verified that for $M = 10^3$, the resulting scalar QNMs are in excellent agreement with those of the classical Schwarzschild black hole \cite{Batic2024EPJC}. Based on this, our focus shifts to the case $M = 1$, where the deviations from the classical Schwarzschild QNM are expected to be the most pronounced. In that regard, we recall that the expansion \eqref{expand} shows that, for $M \gg 1$, the SDG corrections to the geometry in the neighbourhood of the photon sphere are parametrically small, scaling as $1/M^{2}$. Since, in the eikonal limit, the dominant QNMs are determined by the properties of the circular null geodesic in this region, the associated fractional shifts in the QNM frequencies inherit the same scaling, i.e.
\begin{equation}\label{scaling}
\frac{\delta \omega}{\omega_{\rm Sch}} = \mathcal{O}\left(\frac{|\alpha|}{M^{2}}\right) .
\end{equation}
Our numerical results are fully consistent with this behaviour, that is,  for $M = 10^{3}$ the QNM spectrum is indistinguishable, within our quoted precision, from that of the Schwarzschild black hole, whereas for $M = \mathcal{O}(1)$ the same scaling allows $\mathcal{O}(1)$ fractional shifts, which is precisely the regime explored in our numerical tables and figures.

\begin{itemize}
\item
The WKB approximation employed in \cite{Lambiase2023EPJC} fails to capture the presence of purely overdamped modes. This limitation is well known and has also been documented in previous analysis (see, e.g., \cite{Batic2024PRD, Batic2024EPJC}). Moreover, \cite{Lambiase2023EPJC} does not report any overtones and does not address the gravitational perturbation sector, thus limiting the scope of its QNM analysis. The omission of overtones limits the completeness of the spectral profile, while neglecting gravitational modes precludes a full understanding of the spacetime's stability and dynamical response to metric perturbations. Finally, we note that \cite{Lambiase2023EPJC} restricts its analysis to the case $\gamma = 0.5$, without exploring the QNMs corresponding to the theoretically motivated value $\gamma = 9/2$, as proposed in \cite{Bonanno2000PRD}. This omission overlooks an important regime that is more closely tied to the asymptotic safety framework.
\item 
Regarding the morphology of the QNM spectrum, we observe remarkable stability across all examined values of $\gamma$, angular momentum $\ell$, and perturbation sectors. The spectrum consistently follows a characteristic Martini glass distribution (see Figure~\ref{Martini}), underscoring its robustness as a structural feature of the model.
\begin{figure}
    \centering
    \includegraphics[width=0.49\textwidth]{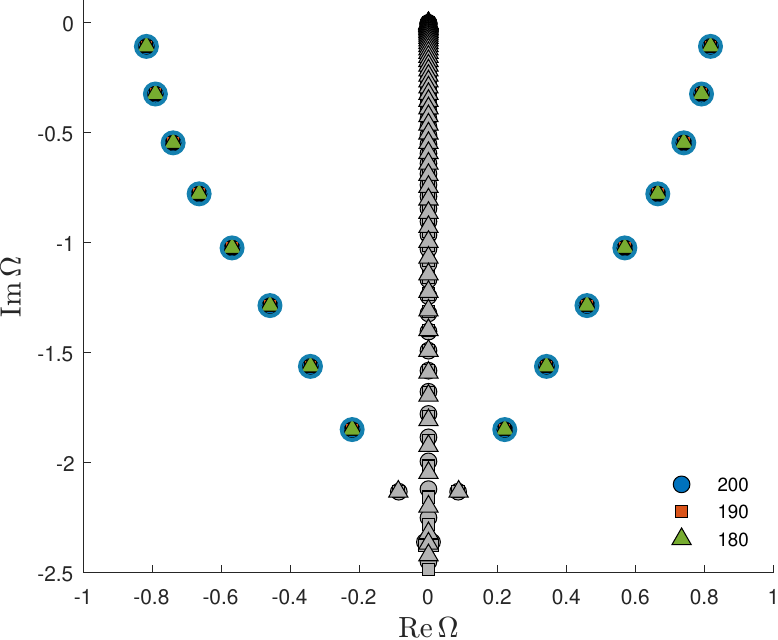}
    \caption{Representative "Martini glass" morphology of the QNM spectrum, a pattern consistently observed across all perturbation sectors. The example shown corresponds to the gravitational case with spin $s = 2$, coupling parameter $\gamma = 9/2$, and angular momentum $\ell = 5$.}
    \label{Martini}
\end{figure}
\item 
For the scalar case ($s = 0$) with $\gamma = 0.5$, the oscillatory QNMs are presented in Table~\ref{scalarg05}. For angular momentum values $\ell \in \{0, 1, \ldots, 9\}$, we list the fundamental QNM along with a selection of overtones. Our results exhibit good agreement with the fundamental modes reported in \cite{Lambiase2023EPJC}, with the exception that the latter was unable to determine the fundamental mode for $\ell=0$. However, while the latter does not explore overtone modes, our SM with 200 Chebyshev polynomials successfully resolves two oscillatory QNMs for $\ell = 0$, increasing to 25 distinct modes for $\ell = 9$. For each value of $\ell$ considered, we detected more than 88 nearly equally spaced overdamped QNMs, detailed in Table~\ref{scalarg05overdamped}. The spacing between successive non-oscillatory modes is denoted by $\Delta\omega$, and we observe that this spacing becomes increasingly stable as $\ell$ increases. Interestingly, for $\ell \in \{0, 2, 4\}$, the gap $\Delta\omega$ between the fundamental overdamped mode and its first overtone appears to be approximately four times larger than the typical spacing. This anomaly reduces to a double gap for $\ell = 7$ and transitions to a triple gap for $\ell \in \{7, 8\}$. Furthermore, for fixed $\ell$, we find that $\Delta\omega$ decreases very gradually with increasing overtone number, reaching a minimum before increasing again. For example, in the case $\ell = 3$, this minimum occurs around the 25th overtone, whereas for $\ell = 5$, it is observed near the 80th overtone. In the scalar case with $\gamma = 9/2$, we detect only a single oscillatory mode for $\ell = 0$, and two such modes for $\ell = 1$ (see Table~\ref{scalarg92}). However, the number of resolved overtones increases rapidly with $\ell$. For instance, at $\ell = 9$, using 200 Chebyshev polynomials, we are able to identify 15 distinct oscillatory overtones. Regarding the overdamped modes, we find that they are present for all values of $\ell$ considered, and their frequencies are nearly equally spaced (see Table~\ref{scalarg92overdamped}). On average, we detect approximately 87 overdamped modes per angular momentum value. Unlike the $\gamma = 0.5$ case, where multiple anomalous gaps were observed, the presence of an enlarged spacing between the fundamental overdamped mode and the first overtone occurs only for higher multipoles, specifically beginning at $\ell = 9$, where the gap reaches approximately twice the typical intermode separation.
\item 
For the electromagnetic perturbations with $\gamma=0.5$, \cite{Lambiase2023EPJC} is able to compute only the fundamental mode for $\ell=1$, whereas we also detect the first two overtones: this number increases to 26 for $\ell=10$ (see Table~\ref{vectorg05}). We further observe that the 6th-order WKB method used in \cite{Lambiase2023EPJC} systematically overestimates the real part of the QNM by approximately 1\% and underestimates the imaginary part by roughly the same amount. Concerning the overdamped modes, we detect more than 86 for each value of $\ell$ considered. For $\ell\in\{7, 9, 10\}$, a double gap is observed between the fundamental overdamped mode and the first overtone, while for $\ell = 8$ this gap becomes triple (see Table~\ref{vectorg05overdamped}). When transitioning to the case $\gamma = 9/2$, we are able to detect up to 18 overtones for $\ell = 10$. However, due to space constraints, we report only the fundamental QNM and the first three overtones for each $\ell$ (see Table~\ref{vectorg92}). Interestingly, for $\ell \geq 4$, the imaginary part of the fundamental QNM remains constant as $\ell$ increases. In this case as well, we identify several equally spaced overdamped modes (see Table~\ref{vectorg92overdamped}). Last but not least, for $\ell = 8$, a double gap occurs between the first and second overtone, whereas for $\ell = 9$, the double gap appears between the fundamental mode and the first overtone.
\item 
We conclude our analysis by examining the gravitational perturbation sector, which was entirely omitted in \cite{Lambiase2023EPJC} for unspecified reasons. For the case $\gamma = 0.5$, we detect a minimum of four QNMs at $\ell = 2$, increasing to 28 QNMs at $\ell = 11$ (see Table~\ref{tensorg05}). It is worth noticing that, for this value of $\gamma$, no large sets of equally spaced overdamped modes emerge. However, a limited number of overdamped modes is observed, and only for $\ell \geq 8$ (see Table~\ref{tensorg05overdamped}). The situation changes significantly in the case $\gamma = 9/2$, where we detect on average 20 overdamped modes for each value of $\ell$. A striking feature highlighted in Table~\ref{tensorg92overdamped} is the emergence of isolated overdamped modes. For instance, at $\ell = 6$, the fundamental overdamped mode is separated by a gap approximately 50 times larger than the spacing between subsequent overtones, suggesting its interpretation as a distinct or isolated mode. This phenomenon becomes more elaborate for higher $\ell$. At $\ell = 7$, two isolated overdamped modes are identified: the first and second are separated by a triple gap, and the second is followed by a gap 47 times larger than the typical overtone spacing. At $\ell = 8$, one isolated overdamped mode is observed, separated by 42 typical gaps from the first overtone. Similarly, for $\ell = 9$, we find one isolated mode with a 32-gap separation. At $\ell = 10$, three isolated overdamped modes appear, evenly spaced by $\Delta\omega = 0.3878i$, with the third mode followed by a gap approximately 30 times the typical overtone spacing. Finally, for $\ell = 11$, a single isolated overdamped mode is detected, located 25 gaps away from the first overtone. Last but not least, Table~\ref{tensorg92} presents the oscillatory QNMs, where we observe a very slow increase in the imaginary part of the fundamental mode as $\ell$ increases.
\end{itemize}

\begin{table}
\centering
\caption{QNMs for scalar perturbations of a black hole in Scale Dependent Gravity with parameters $M = 1$, $\alpha = -\frac{41}{10\pi}$, $\gamma = 0.5$, and angular momentum $\ell \in \{0, 1, 2, \ldots, 9\}$. The event horizon is located at $r_h=2.598800026$. The results are obtained using our SM with $200$ Chebyshev polynomials and $200-$ digit precision. Here, $\omega$ denotes the dimensionless QNM frequency, and $N$ the overtone number. 'N/A' indicates data not available; 'SM' refers to the Spectral Method; and 'WKB' denotes the Padé-averaged sixth-order WKB approximation used in \cite{Lambiase2023EPJC}.}
\label{scalarg05}
\vspace*{1em}
\begin{tabular}{||c|c|c|c|c|c|c|c|c|c|c||}
\hline\hline
$\ell$ & $N$ & $\omega$ \cite{Lambiase2023EPJC} (WKB) & $\omega$ (SM)  & $\ell$ & $N$ & $\omega$ \cite{Lambiase2023EPJC} (WKB)  & $\omega$ (SM) \\ [0.5ex]
\hline\hline
$0$    & $0$ & N/A                                    & $0.0882-0.1119i$ & $5$    & $0$ & $0.94302-0.0998188i$                    & $0.9332-0.1027i$\\
       & $1$ & N/A                                    & $0.0493-0.3904i$ &        & $1$ & N/A                                     & $0.9173-0.3098i$\\
       & $2$ & N/A                                    & N/A              &        & $2$ & N/A                                     & $0.8862-0.5221i$\\
       & $3$ & N/A                                    & N/A              &        & $3$ & N/A                                     & $0.8416-0.7430i$\\
$1$    & $0$ & $0.258266-0.101353i$                   & $0.2543-0.1046i$ & $6$    & $0$ & $1.11437-0.099777i$                     & $1.1028-0.1026i$\\
       & $1$ & N/A                                    & $0.2055-0.3357i$ &        & $1$ & N/A                                     & $1.0893-0.3091i$\\
       & $2$ & N/A                                    & $0.1451-0.6091i$ &        & $2$ & N/A                                     & $1.0628-0.5193i$\\
       & $3$ & N/A                                    & $0.0945-0.9041i$ &        & $3$ & N/A                                     & $1.0241-0.7357i$\\
$2$    & $0$ & $0.429179-0.100341i$                   & $0.4242-0.1035i$ & $7$    & $0$ & $1.28574-0.0997505i$                    & $1.2725-0.1026i$\\
       & $1$ & N/A                                    & $0.3905-0.3186i$ &        & $1$ & N/A                                     & $1.2607-0.3087i$\\
       & $2$ & N/A                                    & $0.3328-0.5569i$ &        & $2$ & N/A                                     & $1.2376-0.5176i$\\
       & $3$ & N/A                                    & $0.2693-0.8220i$ &        & $3$ & N/A                                     & $1.2036-0.7311i$\\
$3$    & $0$ & $0.600393-0.100027i$                   & $0.5939-0.1030i$ & $8$    & $0$ & $1.45711-0.0997326i$                    & $1.4421-0.1026i$\\
       & $1$ & N/A                                    & $0.5692-0.3133i$ &        & $1$ & N/A                                     & $1.4317-0.3084i$\\
       & $2$ & N/A                                    & $0.5230-0.5362i$ &        & $2$ & N/A                                     & $1.4112-0.5164i$\\
       & $3$ & N/A                                    & $0.4631-0.7780i$ &        & $3$ & N/A                                     & $1.3810-0.7280i$\\
$4$    & $0$ & $0.771688-0.0998901i$                  & $0.7636-0.1028i$ & $9$    & $0$ & N/A                                     & $1.6117-0.1025i$\\
       & $1$ & N/A                                    & $0.7442-0.3110i$ &        & $1$ & N/A                                     & $1.6025-0.3082i$\\
       & $2$ & N/A                                    & $0.7068-0.5269i$ &        & $2$ & N/A                                     & $1.5841-0.5156i$\\
       & $3$ & N/A                                    & $0.6549-0.7553i$ &        & $3$ & N/A                                     & $1.5569-0.7258i$\\[1ex]
\hline\hline 
\end{tabular}
\end{table}

\begin{table}
\centering
\caption{Purely imaginary QNMs for scalar perturbations of a black hole in Scale Dependent Gravity with parameters $M = 1$, $\alpha = -\frac{41}{10\pi}$, $\gamma = 0.5$, and angular momentum $\ell \in \{0, 1, 2, \ldots, 9\}$. The corresponding results are obtained through our SM, utilising $200$ polynomials with a precision of $200$ digits. In this context, $\omega$ and $N$ represent the dimensionless frequency and the corresponding overtone, respectively, while $\Delta\omega=\omega_N-\omega_{N+1}$. The notation 'SM' stands for Spectral Method.}
\label{scalarg05overdamped}
\vspace*{1em}
\begin{tabular}{||c|c|c|c|c|c|c|c|c||}
\hline\hline
$\ell$ & $N$ & $\omega$ (SM)               &$\Delta\omega$ & $\ell$ & $N$ & $\omega$ (SM)  &$\Delta\omega$\\ [0.5ex]
\hline\hline
$0$    & $0$ & $0.0000-2.8772i$            & $1.2606i$           & $5$    & $0$ & $0.0000-7.9160i$            &$0.3180i$\\
       & $1$ & $0.0000-4.1378i$            & $0.3156i$           &        & $1$ & $0.0000-8.2340i$            &$0.3178i$\\
       & $2$ & $0.0000-4.4534i$            & $0.3159i$           &        & $2$ & $0.0000-8.5517i$            &$0.3177i$\\
       & $3$ & $0.0000-4.7693i$            & $0.3160i$           &        & $3$ & $0.0000-8.8694i$            &$0.3176i$\\
       & $4$ & $0.0000-5.0853i$            & $0.3161i$           &        & $4$ & $0.0000-9.1870i$            &$0.3175i$\\
       & $5$ & $0.0000-5.4013i$            & $0.3162i$           &        & $5$ & $0.0000-9.5045i$            &$0.3174i$\\
$1$    & $0$ & $0.0000-4.4507i$            & $0.3163i$           & $6$    & $0$ & $0.0000-8.8619i$            &$0.3181i$\\
       & $1$ & $0.0000-4.7670i$            & $0.3163i$           &        & $1$ & $0.0000-9.1800i$            &$0.3179i$\\
       & $2$ & $0.0000-5.0833i$            & $0.3164i$           &        & $2$ & $0.0000-9.4979i$            &$0.3179i$\\
       & $3$ & $0.0000-5.3997i$            & $0.3164i$           &        & $3$ & $0.0000-9.8158i$            &$0.3178i$\\
       & $4$ & $0.0000-5.7161i$            & $0.3164i$           &        & $4$ & $0.0000-10.1335i$           &$0.3177i$\\
       & $5$ & $0.0000-6.0325i$            & $0.3164i$           &        & $5$ & $0.0000-10.4512i$           &$0.3176i$\\
$2$    & $0$ & $0.0000-3.8094i$            & $1.2698i$           & $7$    & $0$ & $0.0000-9.8074i$            &$0.6365i$\\
       & $1$ & $0.0000-5.0793i$            & $0.3169i$           &        & $1$ & $0.0000-10.4439i$           &$0.3181i$\\
       & $2$ & $0.0000-5.3962i$            & $0.3168i$           &        & $2$ & $0.0000-10.7620i$           &$0.3179i$\\
       & $3$ & $0.0000-5.7130i$            & $0.3168i$           &        & $3$ & $0.0000-11.0799i$           &$0.3179i$\\
       & $4$ & $0.0000-6.0298i$            & $0.3168i$           &        & $4$ & $0.0000-11.3978i$           &$0.3178i$\\
       & $5$ & $0.0000-6.3465i$            & $0.3168i$           &        & $5$ & $0.0000-11.7156i$           &$0.3177i$\\      
$3$    & $0$ & $0.0000-5.7072i$            & $0.3174i$           & $8$    & $0$ & $0.0000-10.7533i$           &$0.9548i$\\
       & $1$ & $0.0000-6.0246i$            & $0.3173i$           &        & $1$ & $0.0000-11.7081i$           &$0.3181i$\\
       & $2$ & $0.0000-6.3419i$            & $0.3173i$           &        & $2$ & $0.0000-12.0262i$           &$0.3180i$\\
       & $3$ & $0.0000-6.6592i$            & $0.3172i$           &        & $3$ & $0.0000-12.3442i$           &$0.3179i$\\
       & $4$ & $0.0000-6.9764i$            & $0.3171i$           &        & $4$ & $0.0000-12.6621i$           &$0.3179i$\\
       & $5$ & $0.0000-7.2934i$            & $0.3170i$           &        & $5$ & $0.0000-12.9800i$           &$0.3178i$\\      
$4$    & $0$ & $0.0000-5.6957i$            & $1.2747i$           & $9$    & $0$ & $0.0000-11.6985i$           &$0.9556i$\\
       & $1$ & $0.0000-6.9704i$            & $0.3175i$           &        & $1$ & $0.0000-12.6541i$           &$0.3182i$\\
       & $2$ & $0.0000-7.2879i$            & $0.3176i$           &        & $2$ & $0.0000-12.9723i$           &$0.3181i$\\
       & $3$ & $0.0000-7.6055i$            & $0.3174i$           &        & $3$ & $0.0000-13.2905i$           &$0.3181i$\\
       & $4$ & $0.0000-7.9229i$            & $0.3174i$           &        & $4$ & $0.0000-13.6085i$           &$0.3180i$\\
       & $5$ & $0.0000-8.2403i$            & $0.3173i$           &        & $5$ & $0.0000-13.9265i$           &$0.3179i$\\[1ex]
 \hline\hline 
 \end{tabular}
\end{table}

\begin{table}
\centering
\caption{QNMs for vector perturbations of a black hole in Scale Dependent Gravity with parameters $M = 1$, $\alpha = -\frac{41}{10\pi}$, $\gamma = 0.5$, and angular momentum $\ell \in \{1, 2,\ldots, 10\}$. The event horizon is located at $r_h=2.598800026$. The results are obtained using our SM with $200$ Chebyshev polynomials and $200-$digit precision. Here, $\omega$ denotes the dimensionless QNM frequency, and $N$ the overtone number. 'N/A' indicates data not available; 'SM' refers to the Spectral Method; and 'WKB' denotes the Padé-averaged sixth-order WKB approximation used in \cite{Lambiase2023EPJC}.}
\label{vectorg05}
\vspace*{1em}
\begin{tabular}{||c|c|c|c|c|c|c|c|c|c|c||}
\hline\hline
$\ell$ & $N$ & $\omega$ \cite{Lambiase2023EPJC} (WKB) & $\omega$ (SM)  & $\ell$ & $N$ & $\omega$ \cite{Lambiase2023EPJC} (WKB)  & $\omega$ (SM) \\ [0.5ex]
\hline\hline
$1$    & $0$ & $0.21244-0.0937687i$                   & $0.2073-0.0962i$              & $6$    & $0$ & $1.10435-0.0994459i$            & $1.0927-0.1023i$\\
       & $1$ & N/A                                    & $0.1513-0.3172i$              &        & $1$ & N/A                             & $1.0790-0.3081i$\\
       & $2$ & N/A                                    & $0.0856-0.5955i$              &        & $2$ & N/A                             & $1.0522-0.5177i$\\
       & $3$ & N/A                                    & N/A                           &        & $3$ & N/A                             & $1.0931-0.7335i$\\
$2$    & $0$ & $0.402615-0.0979465i$                  & $0.3971-0.1010i$              & $7$    & $0$ & $1.27706-0.0995025i$            & $1.2637-0.1023i$\\
       & $1$ & N/A                                    & $0.3612-0.3117i$              &        & $1$ & N/A                             & $1.2519-0.3079i$\\
       & $2$ & N/A                                    & $0.2993-0.5476i$              &        & $2$ & N/A                             & $1.2285-0.5164i$\\
       & $3$ & N/A                                    & $0.2311-0.8125i$              &        & $3$ & N/A                             & $1.1943-0.7295i$\\
$3$    & $0$ & $0.581623-0.0988517i$                  & $0.5748-0.1018i$              & $8$    & $0$ & $1.44946-0.0995400i$            & $1.4343-0.1024i$\\
       & $1$ & N/A                                    & $0.5492-0.3098i$              &        & $1$ & N/A                             & $1.4239-0.3078i$\\
       & $2$ & N/A                                    & $0.5013-0.5308i$              &        & $2$ & N/A                             & $1.4033-0.5154i$\\
       & $3$ & N/A                                    & $0.4390-0.7716i$              &        & $3$ & N/A                             & $1.3729-0.7267i$\\
$4$    & $0$ & $0.757157-0.0991906i$                  & $0.7488-0.1021i$              & $9$    & $0$ & N/A                             & $1.6048-0.1024i$\\
       & $1$ & N/A                                    & $0.7290-0.3089i$              &        & $1$ & N/A                             & $1.5955-0.3077i$\\
       & $2$ & N/A                                    & $0.6908-0.5236i$              &        & $2$ & N/A                             & $1.5770-0.5148i$\\
       & $3$ & N/A                                    & $0.6377-0.7510i$              &        & $3$ & N/A                             & $1.5497-0.7248i$\\
$5$    & $0$ & $0.931159-0.0993542i$                  & $0.9212-0.1022i$              & $10$   & $0$ & N/A                             & $1.7751-0.1024i$\\
       & $1$ & N/A                                    & $0.9050-0.3084i$              &        & $1$ & N/A                             & $1.7667-0.3076i$\\
       & $2$ & N/A                                    & $0.8734-0.5199i$              &        & $2$ & N/A                             & $1.7499-0.5143i$\\
       & $3$ & N/A                                    & $0.8282-0.7400i$              &        & $3$ & N/A                             & $1.7251-0.7234i$\\[1ex]
\hline\hline 
\end{tabular}
\end{table}

\begin{table}
\centering
\caption{Purely imaginary QNMs for vector perturbations of a black hole in Scale Dependent Gravity with parameters $M = 1$, $\alpha = -\frac{41}{10\pi}$, $\gamma = 0.5$, and angular momentum $\ell \in \{1,2,\ldots,10\}$. The corresponding results are obtained through our SM, utilising $200$ polynomials with a precision of $200$ digits. In this context, $\omega$ and $N$ represent the dimensionless frequency and the corresponding overtone, respectively, while $\Delta\omega = \omega_N - \omega_{N+1}$. The notation 'SM' stands for Spectral Method.}
\label{vectorg05overdamped}
\vspace*{1em}
\begin{tabular}{||c|c|c|c|c|c|c|c|c||}
\hline\hline
$\ell$ & $N$ & $\omega$ (SM)               &$\Delta\omega$ & $\ell$ & $N$ & $\omega$ (SM)  &$\Delta\omega$\\ [0.5ex]
\hline\hline
$1$    & $0$ & $0.0000-3.1677i$                         & $0.3172i$           & $6$    & $0$ & $0.0000-8.8568i$            &$0.3181i$\\
       & $1$ & $0.0000-3.4849i$                         & $0.3171i$           &        & $1$ & $0.0000-9.1749i$            &$0.3180i$\\
       & $2$ & $0.0000-3.8020i$                         & $0.3171i$           &        & $2$ & $0.0000-9.4929i$            &$0.3179i$\\
       & $3$ & $0.0000-4.1191i$                         & $0.3170i$           &        & $3$ & $0.0000-9.8108i$            &$0.3178i$\\
       & $4$ & $0.0000-4.4361i$                         & $0.3170i$           &        & $4$ & $0.0000-10.1286i$           &$0.3177i$\\
       & $5$ & $0.0000-4.7531i$                         & $0.3170i$           &        & $5$ & $0.0000-10.4464i$           &$0.3177i$\\
$2$    & $0$ & $0.0000-4.1157i$                         & $0.3176i$           & $7$    & $0$ & $0.0000-9.8030i$            &$0.6365i$\\
       & $1$ & $0.0000-4.4333i$                         & $0.3174i$           &        & $1$ & $0.0000-10.4395i$           &$0.3181i$\\
       & $2$ & $0.0000-4.7507i$                         & $0.3173i$           &        & $2$ & $0.0000-10.7576i$           &$0.3180i$\\
       & $3$ & $0.0000-5.0680i$                         & $0.3172i$           &        & $3$ & $0.0000-11.0756i$           &$0.3179i$\\
       & $4$ & $0.0000-5.3852i$                         & $0.3172i$           &        & $4$ & $0.0000-11.3935i$           &$0.3178i$\\
       & $5$ & $0.0000-5.7024i$                         & $0.3171i$           &        & $5$ & $0.0000-11.7113i$           &$0.3177i$\\
$3$    & $0$ & $0.0000-5.0628i$                         & $0.3178i$           & $8$    & $0$ & $0.0000-10.7493i$           &$0.9549i$\\
       & $1$ & $0.0000-5.3806i$                         & $0.3177i$           &        & $1$ & $0.0000-11.7042i$           &$0.3181i$\\
       & $2$ & $0.0000-5.6983i$                         & $0.3176i$           &        & $2$ & $0.0000-12.0223i$           &$0.3180i$\\
       & $3$ & $0.0000-6.0159i$                         & $0.3175i$           &        & $3$ & $0.0000-12.3404i$           &$0.3180i$\\
       & $4$ & $0.0000-6.3334i$                         & $0.3174i$           &        & $4$ & $0.0000-12.6583i$           &$0.3179i$\\
       & $5$ & $0.0000-6.6509i$                         & $0.3173i$           &        & $5$ & $0.0000-12.9762i$           &$0.3178i$\\      
$4$    & $0$ & $0.0000-5.6900i$                         & $0.3189i$           & $9$    & $0$ & $0.0000-11.6950i$           &$0.3187i$\\
       & $1$ & $0.0000-6.0089i$                         & $0.3185i$           &        & $1$ & $0.0000-12.0137i$           &$0.6369i$\\
       & $2$ & $0.0000-6.3273i$                         & $0.3181i$           &        & $2$ & $0.0000-12.6506i$           &$0.3183i$\\
       & $3$ & $0.0000-6.6454i$                         & $0.3179i$           &        & $3$ & $0.0000-12.9688i$           &$0.3182i$\\
       & $4$ & $0.0000-6.9633i$                         & $0.3177i$           &        & $4$ & $0.0000-13.2870i$           &$0.3181i$\\
       & $5$ & $0.0000-7.2810i$                         & $0.3177i$           &        & $5$ & $0.0000-13.6051i$           &$0.3180i$\\      
$5$    & $0$ & $0.0000-7.9101i$                         & $0.3180i$           & $10$   & $0$ & $0.0000-12.6412i$           &$0.3186i$\\
       & $1$ & $0.0000-8.2281i$                         & $0.3178i$           &        & $1$ & $0.0000-12.9598i$           &$0.3186i$\\
       & $2$ & $0.0000-8.5459i$                         & $0.3178i$           &        & $2$ & $0.0000-13.2784i$           &$0.6368i$\\
       & $3$ & $0.0000-8.8637i$                         & $0.3177i$           &        & $3$ & $0.0000-13.9152i$           &$0.3183i$\\
       & $4$ & $0.0000-9.1814i$                         & $0.3176i$           &        & $4$ & $0.0000-14.2335i$           &$0.3182i$\\
       & $5$ & $0.0000-9.4989i$                         & $0.3175i$           &        & $5$ & $0.0000-14.5517i$           &$0.3181i$\\[1ex]
 \hline\hline 
 \end{tabular}
\end{table}

\begin{table}
\centering
\caption{QNMs for tensor perturbations of a black hole in Scale Dependent Gravity with parameters $M = 1$, $\alpha = -\frac{41}{10\pi}$, $\gamma = 0.5$, and angular momentum $\ell \in \{2, 3,\ldots, 11\}$. The event horizon is located at $r_h=2.598800026$. The results are obtained using our SM with $200$ Chebyshev polynomials and $200-$digit precision. Here, $\omega$ denotes the dimensionless QNM frequency, and $N$ the overtone number. 'N/A' indicates data not available, and 'SM' refers to the Spectral Method.}
\label{tensorg05}
\vspace*{1em}
\begin{tabular}{||c|c|c|c|c|c|c|c|c|c|c||}
\hline\hline
$\ell$ & $N$ & $\omega$ (SM)  & $\ell$ & $N$ & $\omega$ (SM) \\ [0.5ex]
\hline\hline
$2$    & $0$ & $0.3137-0.0976i$            & $7$    & $0$ & $1.2372-0.1016i$\\
       & $1$ & $0.2549-0.3050i$            &        & $1$ & $1.2251-0.3058i$\\
       & $2$ & $0.1404-0.5902i$            &        & $2$ & $1.2012-0.5129i$\\
       & $3$ & $0.1011-0.9131i$            &        & $3$ & $1.1660-0.7247i$\\
$3$    & $0$ & $0.5170-0.0990i$            & $8$    & $0$ & $1.4111-0.1018i$\\
       & $1$ & $0.4866-0.3016i$            &        & $1$ & $1.4004-0.3061i$\\
       & $2$ & $0.4288-0.5182i$            &        & $2$ & $1.3794-0.5127i$\\
       & $3$ & $0.3510-0.7567i$            &        & $3$ & $1.3483-0.7229i$\\
$4$    & $0$ & $0.7043-0.1003i$            & $9$    & $0$ & $1.5840-0.1019i$\\
       & $1$ & $0.6827-0.3034i$            &        & $1$ & $1.5745-0.3064i$\\
       & $2$ & $0.6409-0.5148i$            &        & $2$ & $1.5558-0.5126i$\\
       & $3$ & $0.5822-0.7396i$            &        & $3$ & $1.5280-0.7217i$\\
$5$    & $0$ & $0.8850-0.1009i$            & $10$   & $0$ & $1.7563-0.1020i$\\
       & $1$ & $0.8679-0.3046i$            &        & $1$ & $1.7478-0.3065i$\\
       & $2$ & $0.8346-0.5136i$            &        & $2$ & $1.7308-0.5125i$\\
       & $3$ & $0.7867-0.7317i$            &        & $3$ & $1.7056-0.7209i$\\
$6$    & $0$ & $1.0621-0.1014i$            & $11$   & $0$ & $1.9281-0.1021i$\\
       & $1$ & $1.0480-0.3053i$            &        & $1$ & $1.9203-0.3067i$\\
       & $2$ & $1.0201-0.5131i$            &        & $2$ & $1.9049-0.5125i$\\
       & $3$ & $0.9796-0.7273i$            &        & $3$ & $1.8819-0.7202i$\\[1ex]
\hline\hline 
\end{tabular}
\end{table}

\begin{table}
\centering
\caption{Purely imaginary QNMs for tensor perturbations of a black hole in Scale Dependent Gravity with parameters $M = 1$, $\alpha = -\frac{41}{10\pi}$, $\gamma = 0.5$, and angular momentum $\ell \in \{8,9,10,11\}$. The corresponding results are obtained through our SM, utilising $200$ polynomials with a precision of $200$ digits. In this context, $\omega$ and $N$ represent the dimensionless frequency and the corresponding overtone, respectively, while $\Delta\omega=\omega_N-\omega_{N+1}$. The notation 'SM' stands for Spectral Method.}
\label{tensorg05overdamped}
\vspace*{1em}
\begin{tabular}{||c|c|c|c|c|c|c|c|c||}
\hline\hline
$\ell$ & $N$ & $\omega$ (SM)               &$\Delta\omega$ & $\ell$ & $N$ & $\omega$ (SM)      &$\Delta\omega$\\ [0.5ex]
\hline\hline
$8$    & $0$ & $0.0000-11.7252i$           & $0.9536i$     & $10$   & $0$ & $0.0000-22.1879i$  &-\\
       & $1$ & $0.0000-12.6788i$           & $0.3174i$     &        & $1$ & N/A                &-\\
       & $2$ & $0.0000-12.9962i$           & -             &        & $2$ & N/A                &-\\
       & $3$ & N/A                         & -             &        & $3$ & N/A                &-\\
$9$    & $0$ & $0.0000-12.9880i$           & $1.2714i$     & $11$   & $0$ & $0.0000-21.2321i$  &$0.6345i$\\
       & $1$ & $0.0000-14.2594i$           & $0.3177i$     &        & $1$ & $0.0000-21.8666i$  &$0.3173i$\\
       & $2$ & $0.0000-14.5771i$           & -             &        & $2$ & $0.0000-22.1839i$  &$0.3173i$\\
       & $3$ & N/A                         & -             &        & $3$ & $0.0000-22.5012i$  &-\\[1ex]
 \hline\hline 
 \end{tabular}
\end{table}

\begin{table}
\centering
\caption{QNMs for scalar perturbations of a black hole in Scale Dependent Gravity with parameters $M = 1$, $\alpha = -\frac{41}{10\pi}$, $\gamma = 9/2$, and angular momentum $\ell \in \{0, 1,\ldots, 8\}$. The event horizon is located at $r_h=3.055929964$. The results are obtained using our SM with $200$ Chebyshev polynomials and $200-$digit precision. Here, $\omega$ denotes the dimensionless QNM frequency, and $N$ the overtone number. 'N/A' indicates data not available, and 'SM' refers to the Spectral Method.}
\label{scalarg92}
\vspace*{1em}
\begin{tabular}{||c|c|c|c|c|c|c|c|c|c|c||}
\hline\hline
$\ell$ & $N$ & $\omega$ (SM)  & $\ell$ & $N$ & $\omega$ (SM) \\ [0.5ex]
\hline\hline
$0$    & $0$ & $0.0695-0.1134i$  & $5$ & $0$ & $0.8657-0.1092i$\\
       & $1$ & N/A            &        & $1$ & $0.8417-0.3296i$\\
       & $2$ & N/A            &        & $2$ & $0.7944-0.5555i$\\
       & $3$ & N/A            &        & $3$ & $0.7258-0.7914i$\\
$1$    & $0$ & $0.2299-0.1106i$  & $6$ & $0$ & $1.0235-0.1091i$\\
       & $1$ & $0.1529-0.3634i$  &     & $1$ & $1.0032-0.3286i$\\
       & $2$ & N/A            &        & $2$ & $0.9632-0.5522i$\\
       & $3$ & N/A            &        & $3$ & $0.9043-0.7828i$\\
$2$    & $0$ & $0.3906-0.1103i$  & $7$    & $0$ & $1.1812-0.1090i$\\
       & $1$ & $0.3382-0.3403i$            &        & $1$ & $1.1637-0.3280i$\\
       & $2$ & $0.2437-0.5968i$            &        & $2$ & $1.1290-0.5501i$\\
       & $3$ & $0.1384-0.8771i$            &        & $3$ & $1.0776-0.7774i$\\
$3$    & $0$ & $0.5495-0.1098i$            & $8$    & $0$ & $1.3389-0.1089i$\\
       & $1$ & $0.5116-0.3339i$            &        & $1$ & $1.3235-0.3276i$\\
       & $2$ & $0.4393-0.5721i$            &        & $2$ & $1.2928-0.5486i$\\
       & $3$ & $0.3434-0.8327i$            &        & $3$ & $1.2473-0.7737i$\\
$4$    & $0$ & $0.7077-0.1095i$            & $9$    & $0$ & $1.4965-0.1089i$\\
       & $1$ & $0.6783-0.3311i$            &        & $1$ & $1.4828-0.3273i$\\
       & $2$ & $0.6209-0.5612i$            &        & $2$ & $1.4554-0.5476i$\\
       & $3$ & $0.5397-0.8060i$            &        & $3$ & $1.4145-0.7712i$\\[1ex]
\hline\hline 
\end{tabular}
\end{table}

\begin{table}
\centering
\caption{Purely imaginary QNMs for scalar perturbations of a black hole in Scale Dependent Gravity with parameters $M = 1$, $\alpha = -\frac{41}{10\pi}$, $\gamma = 9/2$, and angular momentum $\ell \in \{0,1,\ldots,8\}$. The corresponding results are obtained through our SM, utilising $200$ polynomials with a precision of $200$ digits. In this context, $\omega$ and $N$ represent the dimensionless frequency and the corresponding overtone, respectively, while $\Delta\omega=\omega_N-\omega_{N+1}$. The notation 'SM' stands for Spectral Method.}
\label{scalarg92overdamped}
\vspace*{1em}
\begin{tabular}{||c|c|c|c|c|c|c|c|c||}
\hline\hline
$\ell$ & $N$ & $\omega$ (SM)               &$\Delta\omega$ & $\ell$ & $N$ & $\omega$ (SM)  &$\Delta\omega$\\ [0.5ex]
\hline\hline
$0$    & $0$ & $0.0000-1.9562i$                         & $0.3852i$           & $5$    & $0$ & $0.0000-4.2559i$            &$0.3905i$\\
       & $1$ & $0.0000-2.3414i$                         & $0.3859i$           &        & $1$ & $0.0000-4.6464i$            &$0.3898i$\\
       & $2$ & $0.0000-2.7273i$                         & $0.3863i$           &        & $2$ & $0.0000-5.0362i$            &$0.3894i$\\
       & $3$ & $0.0000-3.1136i$                         & $0.3866i$           &        & $3$ & $0.0000-5.4256i$            &$0.3890i$\\
       & $4$ & $0.0000-3.5002i$                         & $0.3869i$           &        & $4$ & $0.0000-5.8146i$            &$0.3888i$\\
       & $5$ & $0.0000-3.8871i$                         & $0.3870i$           &        & $5$ & $0.0000-6.2034i$            &$0.3886i$\\
$1$    & $0$ & $0.0000-2.7253i$                         & $0.3869i$           & $6$    & $0$ & $0.0000-5.0282i$            &$0.3905i$\\
       & $1$ & $0.0000-3.1123i$                         & $0.3870i$           &        & $1$ & $0.0000-5.4187i$            &$0.3900i$\\
       & $2$ & $0.0000-3.4993i$                         & $0.3871i$           &        & $2$ & $0.0000-5.8087i$            &$0.3896i$\\
       & $3$ & $0.0000-3.8864i$                         & $0.3872i$           &        & $3$ & $0.0000-6.1982i$            &$0.3892i$\\
       & $4$ & $0.0000-4.2736i$                         & $0.3873i$           &        & $4$ & $0.0000-6.5875i$            &$0.3890i$\\
       & $5$ & $0.0000-4.6609i$                         & $0.3874i$           &        & $5$ & $0.0000-6.9764i$            &$0.3888i$\\
$2$    & $0$ & $0.0000-2.7209i$                         & $0.3881i$           & $7$    & $0$ & $0.0000-5.4094i$            &$0.3912i$\\
       & $1$ & $0.0000-3.1090i$                         & $0.3878i$           &        & $1$ & $0.0000-5.8006i$            &$0.3906i$\\
       & $2$ & $0.0000-3.4969i$                         & $0.3877i$           &        & $2$ & $0.0000-6.1912i$            &$0.3901i$\\
       & $3$ & $0.0000-3.8845i$                         & $0.3876i$           &        & $3$ & $0.0000-6.5813i$            &$0.3897i$\\
       & $4$ & $0.0000-4.2722i$                         & $0.3876i$           &        & $4$ & $0.0000-6.9710i$            &$0.3894i$\\
       & $5$ & $0.0000-4.6598i$                         & $0.3876i$           &        & $5$ & $0.0000-7.3604i$            &$0.3891i$\\      
$3$    & $0$ & $0.0000-3.4921i$                         & $0.3887i$           & $8$    & $0$ & $0.0000-6.1820i$            &$0.3912i$\\
       & $1$ & $0.0000-3.8807i$                         & $0.3884i$           &        & $1$ & $0.0000-6.5732i$            &$0.3906i$\\
       & $2$ & $0.0000-4.2691i$                         & $0.3882i$           &        & $2$ & $0.0000-6.9638i$            &$0.3902i$\\
       & $3$ & $0.0000-4.6573i$                         & $0.3880i$           &        & $3$ & $0.0000-7.3540i$            &$0.3898i$\\
       & $4$ & $0.0000-5.0454i$                         & $0.3879i$           &        & $4$ & $0.0000-7.7438i$            &$0.3895i$\\
       & $5$ & $0.0000-5.4333i$                         & $0.3879i$           &        & $5$ & $0.0000-8.1333i$            &$0.3893i$\\      
$4$    & $0$ & $0.0000-3.8744i$                         & $0.3896i$           & $9$    & $0$ & $0.0000-6.1704i$            &$0.7843i$\\
       & $1$ & $0.0000-4.2639i$                         & $0.3891i$           &        & $1$ & $0.0000-6.9547i$            &$0.3911i$\\
       & $2$ & $0.0000-4.6530i$                         & $0.3888i$           &        & $2$ & $0.0000-7.3458i$            &$0.3906i$\\
       & $3$ & $0.0000-5.0418i$                         & $0.3885i$           &        & $3$ & $0.0000-7.7365i$            &$0.3902i$\\
       & $4$ & $0.0000-5.4303i$                         & $0.3883i$           &        & $4$ & $0.0000-8.1267i$            &$0.3899i$\\
       & $5$ & $0.0000-5.8186i$                         & $0.3882i$           &        & $5$ & $0.0000-8.5166i$            &$0.3896i$\\[1ex]
 \hline\hline 
 \end{tabular}
\end{table}

\begin{table}
\centering
\caption{QNMs for vector perturbations of a black hole in Scale Dependent Gravity with parameters $M = 1$, $\alpha = -\frac{41}{10\pi}$, $\gamma = 9/2$, and angular momentum $\ell \in \{1, 2,\ldots, 10\}$. The event horizon is located at $r_h=3.055929964$. The results are obtained using our SM with $200$ Chebyshev polynomials and $200-$digit precision. Here, $\omega$ denotes the dimensionless QNM frequency, and $N$ the overtone number. 'N/A' indicates data not available, and 'SM' refers to the Spectral Method.}
\label{vectorg92}
\vspace*{1em}
\begin{tabular}{||c|c|c|c|c|c|c|c|c|c|c||}
\hline\hline
$\ell$ & $N$ & $\omega$ (SM)  & $\ell$ & $N$ & $\omega$ (SM) \\ [0.5ex]
\hline\hline
$1$    & $0$ & $0.1815-0.0982i$            & $6$    & $0$ & $1.0131-0.1087i$\\
       & $1$ & $0.0954-0.3441i$            &        & $1$ & $0.9926-0.3274i$\\
       & $2$ & N/A                         &        & $2$ & $0.9521-0.5502i$\\
       & $3$ & N/A                         &        & $3$ & $0.8925-0.7801i$\\
$2$    & $0$ & $0.3624-0.1069i$            & $7$    & $0$ & $1.1722-0.1087i$\\
       & $1$ & $0.3069-0.3310i$            &        & $1$ & $1.1546-0.3271i$\\
       & $2$ & $0.2040-0.5834i$            &        & $2$ & $1.1195-0.5486i$\\
       & $3$ & $0.0921-0.8560i$            &        & $3$ & $1.0676-0.7753i$\\
$3$    & $0$ & $0.5298-0.1083i$            & $8$    & $0$ & $1.3310-0.1087i$\\
       & $1$ & $0.4906-0.3294i$            &        & $1$ & $1.3155-0.3269i$\\
       & $2$ & $0.4153-0.5648i$            &        & $2$ & $1.2846-0.5475i$\\
       & $3$ & $0.3153-0.8232i$            &        & $3$ & $1.2387-0.7721i$\\
$4$    & $0$ & $0.6926-0.1086i$            & $9$    & $0$ & $1.4895-0.1087i$\\
       & $1$ & $0.6625-0.3285i$            &        & $1$ & $1.4756-0.3268i$\\
       & $2$ & $0.6037-0.5569i$            &        & $2$ & $1.4481-0.5467i$\\
       & $3$ & $0.5203-0.8001i$            &        & $3$ & $1.4070-0.7699i$\\
$5$    & $0$ & $0.8534-0.1087i$            & $10$   & $0$ & $1.6478-0.1087i$\\
       & $1$ & $0.8290-0.3279i$            &        & $1$ & $1.6353-0.3266i$\\
       & $2$ & $0.7810-0.5527i$            &        & $2$ & $1.6104-0.5461i$\\
       & $3$ & $0.7111-0.7875i$            &        & $3$ & $1.5732-0.7683i$\\[1ex]
\hline\hline 
\end{tabular}
\end{table}

\begin{table}
\centering
\caption{Purely imaginary QNMs for vector perturbations of a black hole in Scale Dependent Gravity with parameters $M = 1$, $\alpha = -\frac{41}{10\pi}$, $\gamma = 9/2$, and angular momentum $\ell \in \{1,2,\ldots,9\}$. The corresponding results are obtained through our SM, utilising $200$ polynomials with a precision of $200$ digits. In this context, $\omega$ and $N$ represent the dimensionless frequency and the corresponding overtone, respectively, while $\Delta\omega=\omega_N-\omega_{N+1}$. The notation 'SM' stands for Spectral Method.}
\label{vectorg92overdamped}
\vspace*{1em}
\begin{tabular}{||c|c|c|c|c|c|c|c|c||}
\hline\hline
$\ell$ & $N$ & $\omega$ (SM)               &$\Delta\omega$ & $\ell$ & $N$ & $\omega$ (SM)  &$\Delta\omega$\\ [0.5ex]
\hline\hline
$1$    & $0$ & $0.0000-1.5493i$                         & $0.3887i$           & $6$    & $0$ & $0.0000-5.0243i$            &$0.3906i$\\
       & $1$ & $0.0000-1.9380i$                         & $0.3884i$           &        & $1$ & $0.0000-5.4149i$            &$0.3900i$\\
       & $2$ & $0.0000-2.3263i$                         & $0.3881i$           &        & $2$ & $0.0000-5.8049i$            &$0.3896i$\\
       & $3$ & $0.0000-2.7144i$                         & $0.3880i$           &        & $3$ & $0.0000-6.1945i$            &$0.3893i$\\
       & $4$ & $0.0000-3.1024i$                         & $0.3879i$           &        & $4$ & $0.0000-6.5839i$            &$0.3891i$\\
       & $5$ & $0.0000-3.4903i$                         & $0.3879i$           &        & $5$ & $0.0000-6.9729i$            &$0.3889i$\\
$2$    & $0$ & $0.0000-2.3228i$                         & $0.3891i$           & $7$    & $0$ & $0.0000-5.0143i$            &$0.3917i$\\
       & $1$ & $0.0000-2.7119i$                         & $0.3887i$           &        & $1$ & $0.0000-5.4060i$            &$0.3912i$\\
       & $2$ & $0.0000-3.1006i$                         & $0.3884i$           &        & $2$ & $0.0000-5.7972i$            &$0.3906i$\\
       & $3$ & $0.0000-3.4890i$                         & $0.3882i$           &        & $3$ & $0.0000-6.1879i$            &$0.3901i$\\
       & $4$ & $0.0000-3.8772i$                         & $0.3881i$           &        & $4$ & $0.0000-6.5780i$            &$0.3898i$\\
       & $5$ & $0.0000-4.2653i$                         & $0.3880i$           &        & $5$ & $0.0000-6.9678i$            &$0.3895i$\\
$3$    & $0$ & $0.0000-3.0958i$                         & $0.3895i$           & $8$    & $0$ & $0.0000-5.7872i$            &$0.3918i$\\
       & $1$ & $0.0000-3.4853i$                         & $0.3890i$           &        & $1$ & $0.0000-6.1790i$            &$0.7818i$\\
       & $2$ & $0.0000-3.8744i$                         & $0.3887i$           &        & $2$ & $0.0000-6.9608i$            &$0.3902i$\\
       & $3$ & $0.0000-4.2630i$                         & $0.3885i$           &        & $3$ & $0.0000-7.3511i$            &$0.3899i$\\
       & $4$ & $0.0000-4.6515i$                         & $0.3883i$           &        & $4$ & $0.0000-7.7409i$            &$0.3896i$\\
       & $5$ & $0.0000-5.0398i$                         & $0.3882i$           &        & $5$ & $0.0000-8.1305i$            &$0.3893i$\\      
$4$    & $0$ & $0.0000-3.8689i$                         & $0.3897i$           & $9$    & $0$ & $0.0000-6.1678i$            &$0.7842i$\\
       & $1$ & $0.0000-4.2586i$                         & $0.3893i$           &        & $1$ & $0.0000-6.9520i$            &$0.3911i$\\
       & $2$ & $0.0000-4.6479i$                         & $0.3889i$           &        & $2$ & $0.0000-7.3431i$            &$0.3907i$\\
       & $3$ & $0.0000-5.0368i$                         & $0.3887i$           &        & $3$ & $0.0000-7.7338i$            &$0.3903i$\\
       & $4$ & $0.0000-5.4255i$                         & $0.3885i$           &        & $4$ & $0.0000-8.1241i$            &$0.3900i$\\
       & $5$ & $0.0000-5.8140i$                         & $0.3884i$           &        & $5$ & $0.0000-8.5140i$            &$0.3897i$\\      
$5$    & $0$ & $0.0000-4.2513i$                         & $0.3906i$           & $10$   & $0$ & $0.0000-7.3333i$            &$0.3916i$\\
       & $1$ & $0.0000-4.6419i$                         & $0.3899i$           &        & $1$ & $0.0000-7.7250i$            &$0.3911i$\\
       & $2$ & $0.0000-5.0318i$                         & $0.3895i$           &        & $2$ & $0.0000-8.1161i$            &$0.3907i$\\
       & $3$ & $0.0000-5.4213i$                         & $0.3891i$           &        & $3$ & $0.0000-8.5068i$            &$0.3903i$\\
       & $4$ & $0.0000-5.8104i$                         & $0.3889i$           &        & $4$ & $0.0000-8.8971i$            &$0.3900i$\\
       & $5$ & $0.0000-6.1993i$                         & $0.3887i$           &        & $5$ & $0.0000-9.2871i$            &$0.3898i$\\[1ex]
 \hline\hline 
 \end{tabular}
\end{table}

\begin{table}
\centering
\caption{QNMs for tensor perturbations of a black hole in Scale Dependent Gravity with parameters $M = 1$, $\alpha = -\frac{41}{10\pi}$, $\gamma = 9/2$, and angular momentum $\ell \in \{2, 3,\ldots, 11\}$. The event horizon is located at $r_h=3.055929964$. The results are obtained using our SM with $200$ Chebyshev polynomials and $200-$digit precision. Here, $\omega$ denotes the dimensionless QNM frequency, and $N$ the overtone number. 'N/A' indicates data not available, and 'SM' refers to the Spectral Method.}
\label{tensorg92}
\vspace*{1em}
\begin{tabular}{||c|c|c|c|c|c|c|c|c|c|c||}
\hline\hline
$\ell$ & $N$ & $\omega$ (SM)  & $\ell$ & $N$ & $\omega$ (SM) \\ [0.5ex]
\hline\hline
$2$    & $0$ & $0.2802-0.1082i$            & $7$    & $0$ & $1.1454-0.1079i$\\
       & $1$ & $0.1795-0.3544i$            &        & $1$ & $1.1272-0.3248i$\\
       & $2$ & N/A                         &        & $2$ & $1.0911-0.5447i$\\
       & $3$ & N/A                         &        & $3$ & $1.0374-0.7698i$\\
$3$    & $0$ & $0.4719-0.1063i$            & $8$    & $0$ & $1.3073-0.1081i$\\
       & $1$ & $0.4227-0.3228i$            &        & $1$ & $1.2915-0.3250i$\\
       & $2$ & $0.3239-0.5522i$            &        & $2$ & $1.2599-0.5444i$\\
       & $3$ & $0.1737-0.8083i$            &        & $3$ & $1.2129-0.7677i$\\
$4$    & $0$ & $0.6478-0.1069i$            & $9$    & $0$ & $1.4683-0.1082i$\\
       & $1$ & $0.6142-0.3232i$            &        & $1$ & $1.4542-0.3252i$\\
       & $2$ & $0.5479-0.5474i$            &        & $2$ & $1.4261-0.5442i$\\
       & $3$ & $0.4524-0.7864i$            &        & $3$ & $1.3843-0.7663i$\\
$5$    & $0$ & $0.8168-0.1074i$            & $10$   & $0$ & $1.6287-0.1083i$\\
       & $1$ & $0.7908-0.3239i$            &        & $1$ & $1.6160-0.3254i$\\
       & $2$ & $0.7394-0.5459i$            &        & $2$ & $1.5907-0.5440i$\\
       & $3$ & $0.6641-0.7777i$            &        & $3$ & $1.5530-0.7653i$\\
$6$    & $0$ & $0.9822-0.1077i$            & $11$   & $0$ & $1.7885-0.1084i$\\
       & $1$ & $0.9608-0.3244i$            &        & $1$ & $1.7770-0.3255i$\\
       & $2$ & $0.9184-0.5451i$            &        & $2$ & $1.7540-0.5439i$\\
       & $3$ & $0.8559-0.7728i$            &        & $3$ & $1.7196-0.7645i$\\[1ex]
\hline\hline 
\end{tabular}
\end{table}

\begin{table}
\centering
\caption{Purely imaginary QNMs for tensor perturbations of a black hole in Scale Dependent Gravity with parameters $M = 1$, $\alpha = -\frac{41}{10\pi}$, $\gamma = 9/2$, and angular momentum $\ell \in \{2,3,\ldots,11\}$. The event horizon is located at $r_h=3.055929964$. The corresponding results are obtained through our SM, utilising $200$ polynomials with a precision of $200$ digits. In this context, $\omega$ and $N$ represent the dimensionless frequency and the corresponding overtone, respectively, while $\Delta\omega=\omega_N-\omega_{N+1}$. The notation 'SM' stands for Spectral Method.}
\label{tensorg92overdamped}
\vspace*{1em}
\begin{tabular}{||c|c|c|c|c|c|c|c|c||}
\hline\hline
$\ell$ & $N$ & $\omega$ (SM)               &$\Delta\omega$ & $\ell$ & $N$ & $\omega$ (SM)  &$\Delta\omega$\\ [0.5ex]
\hline\hline
$2$    & $0$ & $0.0000-23.6833i$                         & $0.3878i$           & $7$    & $0$ & $0.0000-4.2516i$             &$1.1769i$\\
       & $1$ & $0.0000-24.0711i$                         & $0.3876i$           &        & $1$ & $0.0000-5.4285i$             &$18.2498i$\\
       & $2$ & $0.0000-24.4587i$                         & $0.3877i$           &        & $2$ & $0.0000-23.6783i$            &$0.3879i$\\
       & $3$ & $0.0000-24.8464i$                         & $0.3876i$           &        & $3$ & $0.0000-24.0662i$            &$0.3877i$\\
       & $4$ & $0.0000-25.2340i$                         & $0.3877i$           &        & $4$ & $0.0000-24.4539i$            &$0.3878i$\\
       & $5$ & $0.0000-25.6217i$                         & $0.3877i$           &        & $5$ & $0.0000-24.8417i$            &$0.3877i$\\
       & $6$ & $0.0000-26.0094i$                         & $0.3877i$           &        & $6$ & $0.0000-25.2294i$            &$0.3878i$\\
       & $7$ & $0.0000-26.3970i$                         & $0.3877i$           &        & $7$ & $0.0000-25.6172i$            &$0.3878i$\\
$3$    & $0$ & $0.0000-23.6827i$                         & $0.3878i$           & $8$    & $0$ & $0.0000-7.3694i$             &$16.3072i$\\
       & $1$ & $0.0000-24.0706i$                         & $0.3876i$           &        & $1$ & $0.0000-23.6766i$            &$0.3879i$\\
       & $2$ & $0.0000-24.4581i$                         & $0.3877i$           &        & $2$ & $0.0000-24.0645i$            &$0.3877i$\\
       & $3$ & $0.0000-24.8458i$                         & $0.3876i$           &        & $3$ & $0.0000-24.4523i$            &$0.3879i$\\
       & $4$ & $0.0000-25.2335i$                         & $0.3877i$           &        & $4$ & $0.0000-24.8401i$            &$0.3878i$\\
       & $5$ & $0.0000-25.6212i$                         & $0.3877i$           &        & $5$ & $0.0000-25.2279i$            &$0.3878i$\\
       & $6$ & $0.0000-26.0089i$                         & $0.3877i$           &        & $6$ & $0.0000-25.6157i$            &$0.3878i$\\
$4$    & $0$ & $0.0000-23.6819i$                         & $0.3878i$           & $9$    & $0$ & $0.0000-10.8639i$            &$12.4231i$\\
       & $1$ & $0.0000-24.0698i$                         & $0.3876i$           &        & $1$ & $0.0000-23.2870i$            &$0.3877i$\\
       & $2$ & $0.0000-24.4574i$                         & $0.3877i$           &        & $2$ & $0.0000-23.6747i$            &$0.3880i$\\
       & $3$ & $0.0000-24.8451i$                         & $0.3877i$           &        & $3$ & $0.0000-24.0627i$            &$0.3878i$\\
       & $4$ & $0.0000-25.2328i$                         & $0.3877i$           &        & $4$ & $0.0000-24.4505i$            &$0.3879i$\\
       & $5$ & $0.0000-25.6205i$                         & $0.3877i$           &        & $5$ & $0.0000-24.8384i$            &$0.3878i$\\  
       & $6$ & $0.0000-26.0082i$                         & $0.3877i$           &        & $6$ & $0.0000-25.2262i$            &$0.3879i$\\
$5$    & $0$ & $0.0000-23.6809i$                         & $0.3878i$           & $10$   & $0$ & $0.0000-10.4699i$            &$0.3889i$\\
       & $1$ & $0.0000-24.0688i$                         & $0.3876i$           &        & $1$ & $0.0000-10.8588i$            &$0.7773i$\\
       & $2$ & $0.0000-24.4564i$                         & $0.3878i$           &        & $2$ & $0.0000-11.6361i$            &$11.6486i$\\
       & $3$ & $0.0000-24.8442i$                         & $0.3877i$           &        & $3$ & $0.0000-23.2848i$            &$0.3878i$\\
       & $4$ & $0.0000-25.2318i$                         & $0.3877i$           &        & $4$ & $0.0000-23.6726i$            &$0.3880i$\\
       & $5$ & $0.0000-25.6196i$                         & $0.3877i$           &        & $5$ & $0.0000-24.0606i$            &$0.3879i$\\
       & $6$ & $0.0000-26.0073i$                         & $0.3877i$           &        & $6$ & $0.0000-24.4484i$            &$0.3879i$\\
       & $7$ & $0.0000-26.3950i$                         & $0.3877i$           &        & $7$ & $0.0000-24.8364i$            &$0.3879i$\\
       & $8$ & $0.0000-26.7827i$                         & $0.3877i$           &        & $8$ & $0.0000-25.2242i$            &$0.3879i$\\
$6$    & $0$ & $0.0000-4.2676i$                          & $19.4121i$          & $11$   & $0$ & $0.0000-13.1862i$            &$9.7079i$\\
       & $1$ & $0.0000-23.6797i$                         & $0.3879i$           &        & $1$ & $0.0000-22.8942i$            &$0.3881i$\\
       & $2$ & $0.0000-24.0676i$                         & $0.3877i$           &        & $2$ & $0.0000-23.2823i$            &$0.3879i$\\
       & $3$ & $0.0000-24.4552i$                         & $0.3878i$           &        & $3$ & $0.0000-23.6702i$            &$0.3880i$\\
       & $4$ & $0.0000-24.8430i$                         & $0.3877i$           &        & $4$ & $0.0000-24.0582i$            &$0.3879i$\\
       & $5$ & $0.0000-25.2307i$                         & $0.3878i$           &        & $5$ & $0.0000-24.4462i$            &$0.3880i$\\
       & $6$ & $0.0000-25.6185i$                         & $0.3877i$           &        & $6$ & $0.0000-24.8341i$            &$0.3879i$\\[1ex]
 \hline\hline 
 \end{tabular}
\end{table}

\section{Conclusions}

In this first part of our study, we have examined the QNM spectrum of the Planck star geometry that arises from an SDG framework when the running parameter $\alpha$ takes the negative value fixed by EFT matching. In this regime, the running Newton coupling is no longer asymptotically safe, so the model should be viewed as an SDG-inspired effective geometry rather than an explicit realisation of ASG. Starting from the RG-improved Schwarzschild metric, we showed that such a choice of $\alpha$ leads, without additional assumptions, to a black-hole spacetime endowed with a finite-size, high-density core of quantum origin that matches the qualitative expectations for Planck stars proposed in earlier works. At the same time, the geometry is not regular. As discussed in Section III, curvature invariants diverge at a finite radius $r_0 > 0$ inside the event horizon, so the classical Schwarzschild singularity is not removed but relocated to an extended singular core of Planckian density. In this sense, our use of the term Planck star refers to the presence of this finite-size, Planckian-density core in the SDG-improved black-hole spacetime, rather than to a fully non-singular bouncing solution, while the expected causal structure of a black hole is preserved.

We then applied a high-precision SM to compute the QNM spectrum for scalar, vector, and tensor perturbations. This approach, unlike commonly used WKB techniques, allowed us to resolve the entire spectrum, including higher overtones and the overdamped modes that are often inaccessible to semi-analytic approximations. The numerical results were benchmarked against the Schwarzschild limit, where the SDG corrections vanish, confirming that the known QNM frequencies are recovered with excellent accuracy.

Our analysis revealed clear spectral signatures of the Planck star geometry. In particular, we observed quantitative shifts in oscillation frequencies and damping rates in the quantum-corrected regime, as well as subtle structural changes in the overdamped sector. These effects encode direct imprints of the finite-size core and the modified short-distance behaviour of gravity in the SDG framework. Beyond their intrinsic theoretical interest, such QNM deviations could, in principle, inform future gravitational-wave observations aimed at probing strong-field quantum corrections. More broadly, our results emphasise two important points. First of all, SDG provides a consistent, calculable setting in which non standard black hole geometries, including Planck stars, emerge as exact solutions rather than phenomenological ans\"{a}tze. Robust numerical methods, such as the spectral approach used here, are indispensable for accurately characterising the full QNM spectrum in spacetimes inspired by quantum gravity, where WKB approximations may fail or become unreliable. Finally, we emphasise that these spectral features are obtained for a specific phenomenological scale-setting $k(r)$ within SDG. Exploring how the QNM spectrum changes under alternative, physically motivated prescriptions for $k(r)$ is an interesting direction for future work.

On the observational side, we can say that the fundamental QNMs are in principle visible already in the ring-down patterns detected by present GW observatories, such as LIGO, Virgo, and KAGRA (see e.g. \cite{Yi2024, Berti2025} and references therein).  

The challenges for future observational research on QNMs are therefore manifold. Current observational results will likely need refinement and confirmation. However, we know that, in general, even metrics describing non-standard black holes (for example, geometries incorporating quantum effects such as the Planck stars of the present article) usually approximate Schwarzschild or Kerr solutions for large masses and at large distances from the sources. The large-$M$ expansion of the SDG lapse function in Section II shows explicitly that, in the photon sphere region relevant for ringdown, the deviation from the Schwarzschild geometry scales as $|\delta F| \propto |\alpha|/M^{2}$ in Planck units. In the eikonal limit, the dominant QNMs are controlled by this region, so the associated fractional shifts obey the scaling \eqref{scaling}. For a stellar-mass black hole with $M \sim 10^{38}$ in Planck units, this implies $|\delta\omega|/\omega_{\rm Sch} \lesssim 10^{-76}$, far beyond the reach of any present or foreseeable detector. The QNMs potentially observable in the merging of large black holes should therefore be practically indistinguishable from those generated by the standard Schwarzschild or Kerr metrics. These considerations indicate that any direct QNM signal from individual Planck‑mass objects would be extremely challenging to access with present or foreseeable detectors, and that their most realistic signatures are likely to be indirect, for instance, through cumulative imprints on the primordial or stochastic gravitational‑wave background \cite{Sasaki2025}. Our discussion of such prospects is therefore intended as a qualitative exploration of possible observational consequences, rather than as a detailed forecast. The main results of this work should instead be viewed as a theoretical characterisation of the QNM spectrum of a specific SDG Planck-star model. By contrast, our numerical tables focus on the regime $M = \mathcal{O}(1)$, corresponding to mini or micro black holes for which quantum corrections can produce $\mathcal{O}(1)$ fractional shifts. In this regime, the Planck-star metric leads to the deviations from the usual QNMs of Schwarzschild or Kerr geometries that we have computed, and such corrections could, in principle, become observable. The question is: where can we find or observe such micro- or mini-black holes? A straightforward answer contemplates two different production mechanisms. Micro BH (i.e. BH of a few Planck masses) could perhaps be produced in future very large colliders. However, even the most advanced project, the Future Circular Collider (FCC) considered by CERN \cite{FCC}, will reach energies of order $100$ TeV at most. This will still be very far away from the "magical" threshold of the Planck energy $10^{16}$ TeV, where a large production of micro black holes is expected to happen in a full quantum-gravity regime. Unless, of course, the presence of (unexpected) extradimensions wouldn't be able to lower the Planck threshold from $10^{16}$ TeV to the more achievable $100$ TeV. A much more promising source of micro or mini black holes should be the (very) early universe. Here, we can consider the production of Planckian black holes during the inflationary epoch, or perhaps even earlier \cite{Scardigli2010}. However, such micro BHs should have masses of only a few Planck masses, and their Hawking evaporation should have consumed them long ago. Perhaps some echoes of those processes could be detectable in the GW background noise, through the LISA observatory \cite{LISA}. A more interesting source of QNMs within our scope could be mini black holes formed by pressure waves arising from primordial density fluctuations at the end of the inflationary epoch. Those objects are believed to be much more massive, of order $10^{12}$ kg, and they are expected to complete their Hawking evaporation process in the present times. So the observations of the last stage of the evaporation, when the mini BH weighs a few Planck masses, should provide an ideal stage to measure not only QNMs' fundamental tones, but also harmonics, overtones, and more. Future GW observatories projected to this scope include LISA and, more specifically, the Einstein Telescope \cite{ET}. The following years look set to be very interesting!

This work is the first in a planned three-part series. The second paper will extend our analysis to the Bonanno–Reuter black hole, and the third to the Hayward black hole, both for scalar, vector, and tensor perturbations. Together, these studies aim to provide a systematic and unified picture of QNM phenomenology across a representative set of regular and asymptotically safe gravity–inspired black holes, allowing direct comparisons and highlighting observationally relevant differences.

\section*{Code availability}

All analytical computations reported in this manuscript have been rechecked in the computer algebra system \textsc{Maple}. One \textsc{Maple} sheet corresponding to the scalar case is available in the supplementary materials and in the repository below. The discretisation of differential operators \eqref{L0none}-\eqref{L2none} using the Tchebyshev-type SM is equally performed in the \textsc{Maple} computer algebra system. Finally, the numerical solution of the resulting quadratic eigenvalue problem \eqref{eq:eig} is performed in \textsc{Matlab} software using the \texttt{polyeig} function. The linear regression analysis of the spectral gap with respect to angular momentum was also performed in \textsc{Matlab}. All these routines, scripts, and materials are freely accessible at the following repository:

\begin{itemize}
      \item \url{https://github.com/dutykh/ASafeGravity/}
\end{itemize}

\bibliography{QNMS1}

@article{Abbott2016PRL, 
  title={Observation of Gravitational Waves from a Binary Black Hole Merger},
  author={B. P. Abbott and et al.},
  journal={Phys. Rev. Lett.},
  volume={116},
  pages={061102},
  year={2016}
}

@article{Abbott2019,
    author = {Abbott, B. P. and others},
    collaboration = "LIGO Scientific, Virgo",
    title = "{GWTC-1: A Gravitational-Wave Transient Catalog of Compact Binary Mergers Observed by LIGO and Virgo during the First and Second Observing Runs}",
    eprint = "1811.12907",
    archivePrefix = "arXiv",
    primaryClass = "astro-ph.HE",
    reportNumber = "LIGO-P1800307",
    doi = "10.1103/PhysRevX.9.031040",
    journal = {Phys. Rev. X},
    volume = {9},
    number = "3",
    pages = {031040},
    year = {2019}
}

@ARTICLE{Akhundov2008EJTP,
  author = {A. Akhundov and A. Shiekh},
  title = {A Review of Leading Quantum Gravitational Corrections to Newtonian Gravity},
  journal = {EJTP},
  volume = {5},
  number={17},
  pages = {1},
  year = {2008}
}

@article{Ashtekar2005,
    author = "Ashtekar, Abhay",
    editor = "Calcagni, Gianluca and Papantonopoulos, Lefteris and Siopsis, George and Tsamis, Nikos",
    title = "{Gravity and the quantum}",
    eprint = "gr-qc/0410054",
    archivePrefix = "arXiv",
    doi = "10.1088/1367-2630/7/1/198",
    journal = "New J. Phys.",
    volume = "7",
    pages = "198",
    year = "2005"
}

@ARTICLE{Batic2016EPJC,
  author = {P. Bargueño and S. Bravo Medina and M. Nowakowski and D. Batic},
  title = {Newtonian cosmology with a quantum bounce},
  journal = {Eur. Phys. J. C},
  volume = {76},
  pages = {543},
  year = {2016}
}

@ARTICLE{Batic2017EPL,
  author = {P. Bargueño and S. Bravo Medina and M. Nowakowski and D. Batic},
  title = {Quantum-mechanical corrections to the Schwarzschild black-hole metric},
  journal = {EPL},
  volume = {117},
  number={6},
  pages = {6006},
  year = {2017}
}

@article{Batic2019EPJC, 
  title={Perturbing microscopic black holes inspired by noncommutativity},
  author={D. Batic and N. G. Kelkar and M. Nowakowski and K. Redway},
  journal={Eur. Phys. J. C},
  volume={79},
  number={},
  pages={581},
  year={2019},
  publisher={Springer}
}

@article{Batic2024CQG, 
  title={A Unified Spectral Approach for Quasinormal Modes of Morris-Thorne Wormholes},
  author={D. Batic and D. Dutykh},
  journal={Class. Quantum Grav.},
  volume={41},
  pages={215003},
  year={2024}
}

@article{Batic2024EPJC, 
  title={Quasinormal Modes in Noncommutative Schwarzschild Black Holes: A
  Spectral Analysis},
  author={D. Batic and D. Dutykh},
  journal={Eur. Phys. J. C},
  volume={84},
  pages={622},
  year={2024}
}

@article{Batic2024PRD, 
  title={A Unified Spectral Approach for Quasinormal Modes of Lee-Wick Black Holes},
  author={D. Batic and D. Dutykh and B. Loureiro Giacchini},
  journal={Phys. Rev. D},
  volume={110},
  pages={084032},
  year={2024}
}

@article{Batic2025EPJC, 
  title={Instability analysis of massive static phantom wormholes via the spectral method},
  author={D. Batic and D. Dutykh},
  journal={Eur. Phys. J. C},
  volume={85},
  pages={144},
  year={2025}
}

@article{Batic2025CQG, 
  title={A spectral approach for quasinormal frequencies of noncommutative geometry-inspired wormholes},
  author={D. Batic and D. Dutykh and J. J. Beek},
  journal={Class. Quantum Grav.},
  volume={42},
  pages={085003},
  year={2025}
}

@article{Batic2025PRSA, 
  title={Quasi-normal modes of non-commutative geometry-inspired dirty black holes},
  author={D. Batic and D. Dutykh and Z. Ahmed Babou},
  journal={Proc. R. Soc. Lond. A},
  volume={481},
  pages={20250021},
  year={2025}
}

@article{Berti2009CQG, 
  title={Quasinormal modes of black holes and black branes},
  author={E. Berti and V. Cardoso and A. O Starinets},
  journal={Class. Quantum Grav.},
  volume={26},
  pages={163001},
  year={2009}
}

@article{Berti2025,
    author = "Berti, Emanuele and others",
    title = "{Black hole spectroscopy: from theory to experiment}",
    eprint = "2505.23895",
    archivePrefix = "arXiv",
    primaryClass = "gr-qc",
    journal = "arXiv",
    month = "5",
    year = "2025"
}

@ARTICLE{Bonanno2000PRD,
   author ={A. Bonanno and M. Reuter, M.},
   title = {Renormalization group improved black hole spacetimes},
   journal = {Phys. Rev. D},
   volume = {62},
   pages = {043008},
   year = {2000}
}

@article{Bonanno2004,
    author = "Bonanno, Alfio and Reuter, M.",
    title = "{Cosmological perturbations in renormalization group derived cosmologies}",
    eprint = "astro-ph/0210472",
    archivePrefix = "arXiv",
    reportNumber = "MZ-TH-02-16",
    doi = "10.1142/S0218271804003809",
    journal = "Int. J. Mod. Phys. D",
    volume = "13",
    pages = "107--122",
    year = "2004"
}

@article{Bonanno2016,
    author = "Bonanno, Alfio and Koch, Benjamin and Platania, Alessia",
    title = "{Cosmic Censorship in Quantum Einstein Gravity}",
    eprint = "1610.05299",
    archivePrefix = "arXiv",
    primaryClass = "gr-qc",
    doi = "10.1088/1361-6382/aa6788",
    journal = "Class. Quant. Grav.",
    volume = "34",
    number = "9",
    pages = "095012",
    year = "2017"
}

@article{BonannoCasadio,
    author = "Bonanno, Alfio and Casadio, Roberto and Platania, Alessia",
    title = "{Gravitational antiscreening in stellar interiors}",
    eprint = "1910.11393",
    archivePrefix = "arXiv",
    primaryClass = "gr-qc",
    doi = "10.1088/1475-7516/2020/01/022",
    journal = "JCAP",
    volume = "01",
    pages = "022",
    year = "2020"
}

@ARTICLE{Bjerrum2003PRD,
  author = {N. E. J. Bjerrum-Bohr and J. F. Donoghue and B. R. Holstein},
  title = {Quantum corrections to the Schwarzschild and Kerr metrics},
  journal = {Phys. Rev. D},
  volume = {68},
  pages = {084005},
  year = {2003},
  note = {Erratum-ibid: Phys. Rev. D {\bf 71}, 069904 (2005)},
}

@ARTICLE{Bjerrum2003PRDa,
  author = {N. E. J. Bjerrum-Bohr and J. F. Donoghue and B. R. Holstein},
  title = {Quantum gravitational corrections to the nonrelativistic scattering potential of two masses},
  journal = {Phys. Rev. D},
  volume = {67},
  pages = {084033},
  year = {2003},
  note = {Erratum-ibid: Phys. Rev. D {\bf 71}, 069903 (2005)},
}

@ARTICLE{Bjerrum2015PRL,
  author = {N. E. J. Bjerrum-Bohr and J. F. Donoghue and B. R. Holstein and L. Plant$\acute{e}$ and P. Vanhove},
  title = {Bending of Light in Quantum Gravity},
  journal = {Phys. Rev. Lett. },
  volume = {114},
  pages = {061301},
  year = {2015},
}

@article{Cardoso2003,
    author = "Cardoso, Vitor and Lemos, Jose P. S.",
    title = "{Quasinormal modes of the near extremal Schwarzschild-de Sitter black hole}",
    eprint = "gr-qc/0301078",
    archivePrefix = "arXiv",
    doi = "10.1103/PhysRevD.67.084020",
    journal = "Phys. Rev. D",
    volume = "67",
    pages = "084020",
    year = "2003"
}

@article{Connes96,
    author = "Connes, Alain",
    title = "{Gravity coupled with matter and foundation of noncommutative geometry}",
    eprint = "hep-th/9603053",
    archivePrefix = "arXiv",
    doi = "10.1007/BF02506388",
    journal = "Commun. Math. Phys.",
    volume = "182",
    pages = "155--176",
    year = "1996"
}

@article{Contreras2017,
    author = "Contreras, Ernesto and Rinc{\'o}n, {\'A}ngel and Koch, Benjamin and Bargue{\~n}o, Pedro",
    title = "{A regular scale-dependent black hole solution}",
    eprint = "1711.08400",
    archivePrefix = "arXiv",
    primaryClass = "gr-qc",
    doi = "10.1142/S0218271818500323",
    journal = "Int. J. Mod. Phys. D",
    volume = "27",
    number = "03",
    pages = "1850032",
    year = "2017"
}

@article{Contreras2018,
    author = "Contreras, E. and Bargue{\~n}o, P.",
    title = "{Scale--dependent Hayward black hole and the generalized uncertainty principle}",
    eprint = "1809.00785",
    archivePrefix = "arXiv",
    primaryClass = "gr-qc",
    doi = "10.1142/S0217732318501845",
    journal = "Mod. Phys. Lett. A",
    volume = "33",
    number = "32",
    pages = "1850184",
    year = "2018"
}

@article{Donoghue1994,
    author = "Donoghue, John F.",
    title = "{General relativity as an effective field theory: The leading quantum corrections}",
    eprint = "gr-qc/9405057",
    archivePrefix = "arXiv",
    reportNumber = "UMHEP-408",
    doi = "10.1103/PhysRevD.50.3874",
    journal = "Phys. Rev. D",
    volume = "50",
    pages = "3874--3888",
    year = "1994"
}

@article{DonoghueJPG2015,
    author = "Donoghue, John F. and Holstein, Barry R.",
    title = "{Low Energy Theorems of Quantum Gravity from Effective Field Theory}",
    eprint = "1506.00946",
    archivePrefix = "arXiv",
    primaryClass = "gr-qc",
    reportNumber = "ACFI-T15-05",
    doi = "10.1088/0954-3899/42/10/103102",
    journal = "J. Phys. G",
    volume = "42",
    number = "10",
    pages = "103102",
    year = "2015"
}

@article{Donoghue2019,
    author = "Donoghue, John F.",
    title = "{A Critique of the Asymptotic Safety Program}",
    eprint = "1911.02967",
    archivePrefix = "arXiv",
    primaryClass = "hep-th",
    reportNumber = "ACFI-T19-12",
    doi = "10.3389/fphy.2020.00056",
    journal = "Front. in Phys.",
    volume = "8",
    pages = "56",
    year = "2020"
}

@article{Duff1974,
    author = "Duff, M. J.",
    title = "{Quantum corrections to the schwarzschild solution}",
    doi = "10.1103/PhysRevD.9.1837",
    journal = "Phys. Rev. D",
    volume = "9",
    pages = "1837--1839",
    year = "1974"
}

@article{DuttaRoy2022,
    author = "Dutta Roy, Poulami and Kar, Sayan",
    title = "{Generalized Hayward spacetimes: Geometry, matter, and scalar quasinormal modes}",
    eprint = "2206.04505",
    archivePrefix = "arXiv",
    primaryClass = "gr-qc",
    doi = "10.1103/PhysRevD.106.044028",
    journal = "Phys. Rev. D",
    volume = "106",
    number = "4",
    pages = "044028",
    year = "2022"
}

@online{ET,
  title     = {Einstein Telescope},
  author    = {ET},
  year      = 2025,
  url       = {https://www.einstein-telescope.it/en/home-en/},
  urldate   = {2025-09-22}
}

@online{FCC,
  title     = {Future Circular Collider},
  author    = {CERN},
  year      = 2025,
  url       = {https://home.cern/science/accelerators/future-circular-collider},
  urldate   = {2025-09-22}
}

@article{Ferrari2008,
    author = "Ferrari, Valeria and Gualtieri, Leonardo",
    title = "{Quasi-Normal Modes and Gravitational Wave Astronomy}",
    eprint = "0709.0657",
    archivePrefix = "arXiv",
    primaryClass = "gr-qc",
    doi = "10.1007/s10714-007-0585-1",
    journal = "Gen. Rel. Grav.",
    volume = "40",
    pages = "945--970",
    year = "2008"
}

@article{Flachi2013,
    author = "Flachi, Antonino and Lemos, Jos{\'e} P. S.",
    title = "{Quasinormal modes of regular black holes}",
    eprint = "1211.6212",
    archivePrefix = "arXiv",
    primaryClass = "gr-qc",
    doi = "10.1103/PhysRevD.87.024034",
    journal = "Phys. Rev. D",
    volume = "87",
    number = "2",
    pages = "024034",
    year = "2013"
}

@inbook{Franchini2023,
    author = {Franchini, Nicola and V{\"o}lkel, Sebastian H.},
    title = "{Testing General Relativity with Black Hole Quasi-normal Modes}",
    eprint = "2305.01696",
    archivePrefix = "arXiv",
    primaryClass = "gr-qc",
    doi = {10.1007/978-981-97-2871-8_9},
    year = "2024"
}

@article{Frob2022,
    author = {Fr{\"o}b, M. B. and Rein, C. and Verch, R.},
    title = "{Graviton corrections to the Newtonian potential using invariant observables}",
    eprint = "2109.09753",
    archivePrefix = "arXiv",
    primaryClass = "hep-th",
    doi = "10.1007/JHEP01(2022)180",
    journal = "JHEP",
    volume = "01",
    pages = "180",
    year = "2022"
}

@article{Giesler2019PRX, 
  title={Black Hole Ringdown: The Importance of Overtones},
  author={M. Giesler and M. Isi and M. A. Scheel and S. A. Teukolsky},
  journal={Phys. Rev. X},
  volume={9},
  pages={041060},
  year={2019}
}

@article{Giesler25,
    author = "Giesler, Matthew and others",
    title = "{Overtones and nonlinearities in binary black hole ringdowns}",
    eprint = "2411.11269",
    archivePrefix = "arXiv",
    primaryClass = "gr-qc",
    reportNumber = "YITP-24-154, RIKEN-iTHEMS-Report-24",
    doi = "10.1103/PhysRevD.111.084041",
    journal = "Phys. Rev. D",
    volume = "111",
    number = "8",
    pages = "084041",
    year = "2025"
}

@ARTICLE{Hamber1995PLB,
  author = {H. W. Hamber and S. Liu},
  title = {On the Quantum Corrections to the Newtonian Potential},
  journal = {Phys. Lett. B},
  volume = {357},
  pages = {51},
  year = {1995}
}

@article{Hassannejad2025PRD, 
  title={Gravitational Collapse in Scale-Dependent Gravity},
  author={R. Hassannejad and G. Lambiase and F. Scardigli and F. Shojai},
  journal={Phys. Rev. D},
  volume={111},
  pages={064069},
  year={2025}
}

@article{Hendi2020,
    author = "Hendi, S. H. and Sajadi, S. N. and Khademi, M.",
    title = "{Physical properties of a regular rotating black hole: Thermodynamics, stability, and quasinormal modes}",
    eprint = "2006.11575",
    archivePrefix = "arXiv",
    primaryClass = "gr-qc",
    doi = "10.1103/PhysRevD.103.064016",
    journal = "Phys. Rev. D",
    volume = "103",
    number = "6",
    pages = "064016",
    year = "2021"
}

@article{Horava2009,
    author = "Horava, Petr",
    title = "{Quantum Gravity at a Lifshitz Point}",
    eprint = "0901.3775",
    archivePrefix = "arXiv",
    primaryClass = "hep-th",
    doi = "10.1103/PhysRevD.79.084008",
    journal = "Phys. Rev. D",
    volume = "79",
    pages = "084008",
    year = "2009"
}

@article{Jacobson95,
    author = "Jacobson, Ted",
    title = "{Thermodynamics of space-time: The Einstein equation of state}",
    eprint = "gr-qc/9504004",
    archivePrefix = "arXiv",
    reportNumber = "UMDGR-95-114",
    doi = "10.1103/PhysRevLett.75.1260",
    journal = "Phys. Rev. Lett.",
    volume = "75",
    pages = "1260--1263",
    year = "1995"
}

@ARTICLE{Khriplovich2002JETP,
  author = {I. B. Khriplovich and G. G. Kirilin},
  title = {Quantum power correction to the Newton law},
  journal = {J. Exp. Theor. Phys.},
  volume = {95},
  pages = {981},
  year = {2002}
}

@ARTICLE{Khriplovich2004JETP,
  author = {I. B. Khriplovich and G. G. Kirilin},
  title = {Quantum long-range interactions in general relativity},
  journal = {J. Exp. Theor. Phys.},
  volume = {98},
  pages = {1063},
  year = {2004}
}

@ARTICLE{Kiefer,
  author = {C. Kiefer},
  title = {The semiclassical approximation to quantum gravity and its observational consequences},
  journal = {J. Phys. Conf. Ser.},
  volume = {442},
  pages = {012025},
  year = {2013},
}

@article{Koch:2014cqa,
    author = "Koch, Benjamin and Saueressig, Frank",
    title = "{Black holes within Asymptotic Safety}",
    journal = "Int. J. Mod. Phys. A",
    volume = "29",
    number = "8",
    pages = "1430011",
    year = "2014"
}

@article{Koch16,
    author = "Koch, Benjamin and Reyes, Ignacio A. and Rinc{\'o}n, {\'A}ngel",
    title = "{A scale dependent black hole in three-dimensional space{\textendash}time}",
    eprint = "1606.04123",
    archivePrefix = "arXiv",
    primaryClass = "hep-th",
    doi = "10.1088/0264-9381/33/22/225010",
    journal = "Class. Quant. Grav.",
    volume = "33",
    number = "22",
    pages = "225010",
    year = "2016"
}

@article{Konoplya2011RMP, 
  title={Quasinormal modes of black holes: From astrophysics to string theory},
  author={R. A. Konoplya and A. Zhidenko},
  journal={Rev. Mod. Phys.},
  volume={83},
  pages={793},
  year={2011}
}

@article{KonoplyaPLB2020,
    author = "Konoplya, R. A.",
    title = "{Quantum corrected black holes: quasinormal modes, scattering, shadows}",
    eprint = "1912.10582",
    archivePrefix = "arXiv",
    primaryClass = "gr-qc",
    doi = "10.1016/j.physletb.2020.135363",
    journal = "Phys. Lett. B",
    volume = "804",
    pages = "135363",
    year = "2020"
}

@article{KonoplyaJCAP2022,
    author = "Konoplya, R. A. and Zinhailo, A. F. and Kunz, J. and Stuchlik, Z. and Zhidenko, A.",
    title = "{Quasinormal ringing of regular black holes in asymptotically safe gravity: the importance of overtones}",
    eprint = "2206.14714",
    archivePrefix = "arXiv",
    primaryClass = "gr-qc",
    doi = "10.1088/1475-7516/2022/10/091",
    journal = "JCAP",
    volume = "10",
    pages = "091",
    year = "2022"
}

@article{KonoplyaPRD2023,
    author = "Konoplya, R. A. and Stuchlik, Z. and Zhidenko, A. and Zinhailo, A. F.",
    title = "{Quasinormal modes of renormalization group improved Dymnikova regular black holes}",
    eprint = "2303.01987",
    archivePrefix = "arXiv",
    primaryClass = "gr-qc",
    doi = "10.1103/PhysRevD.107.104050",
    journal = "Phys. Rev. D",
    volume = "107",
    number = "10",
    pages = "104050",
    year = "2023"
}

@ARTICLE{Lambiase2022PRD,
   author ={G. Lambiase and F. Scardigli},
   title = {Generalized uncertainty principle and asymptotically safe gravity},
   journal = {Phys. Rev. D},
   volume = {105},
   pages = {124054},
   year = {2022}
}

@ARTICLE{Lambiase2023EPJC,
   author ={G. Lambiase and R. C. Pantig and D. Jyoti Gogoi and A. \"Ovg\"un},
   title = {Investigating the connection between generalized uncertainty principle and asymptotically safe gravity in black hole signatures through shadow and quasinormal modes},
   journal = {Eur. Phys. J. C},
   volume = {83},
   pages = {679},
   year = {2023}
}

@online{LISA,
  title     = {Laser Interferometer Space Antenna},
  author    = {LISA},
  year      = 2025,
  url       = {https://lisa.nasa.gov/},
  urldate   = {2025-09-22}
}

@article{Liu2012,
    author = "Liu, Dao-Jun and Yang, Bin and Zhai, Yong-Jia and Li, Xin-Zhou",
    title = "{Quasinormal modes for asymptotic safe black holes}",
    eprint = "1205.4792",
    archivePrefix = "arXiv",
    primaryClass = "gr-qc",
    doi = "10.1088/0264-9381/29/14/145009",
    journal = "Class. Quant. Grav.",
    volume = "29",
    pages = "145009",
    year = "2012"
}

@article{Liu2018,
    author = "Liu, Lei-Hua and Prokopec, Tomislav and Starobinsky, Alexei A.",
    title = "{Inflation in an effective gravitational model and asymptotic safety}",
    eprint = "1806.05407",
    archivePrefix = "arXiv",
    primaryClass = "gr-qc",
    doi = "10.1103/PhysRevD.98.043505",
    journal = "Phys. Rev. D",
    volume = "98",
    number = "4",
    pages = "043505",
    year = "2018"
}

@article{Malik2024EPL, 
  title={Analytical QNMs of fields of various spin in the Hayward spacetime},
  author={Z. Malik},
  journal={EPL},
  volume={147},
  pages={69001},
  year={2024}
}

@article{Mamani2022EPJC, 
  title={Revisiting the quasinormal modes of the Schwarzschild black hole: Numerical analysis},
  author={L. A. H. Mamani and A. D. D. Masa and L. T. Sanches and V. T. Zanchin},
  journal={Eur. Phys. J. C},
  volume={82},
  pages={897},
  year={2022},
  publisher={Springer}
}

@article{Niedermaier2006,
    author = "Niedermaier, Max and Reuter, Martin",
    title = "{The Asymptotic Safety Scenario in Quantum Gravity}",
    doi = "10.12942/lrr-2006-5",
    journal = "Living Rev. Rel.",
    volume = "9",
    pages = "5--173",
    year = "2006"
}

@article{Olver1994MAA,
  title={Asymptotic expansions of the coefficients in asymptotic series solutions of linear differential equations},
  author={F. W. J. Olver},
  journal={Methods Appl. Anal.},
  volume={1},
  number={},
  pages={1},
  year={1994}
}

@article{Platania2019,
    author = "Platania, Alessia",
    title = "{Dynamical renormalization of black-hole spacetimes}",
    eprint = "1903.10411",
    archivePrefix = "arXiv",
    primaryClass = "gr-qc",
    doi = "10.1140/epjc/s10052-019-6990-2",
    journal = "Eur. Phys. J. C",
    volume = "79",
    number = "6",
    pages = "470",
    year = "2019"
}

@article{Regge1957PR,
  title={Stability of a Schwarzschild Singularity},
  author={T. Regge and J. A. Wheeler},
  journal={Phys. Rev.},
  volume={108},
  number={},
  pages={1063},
  year={1957}
}

@article{Reuter98,
    author = "Reuter, M.",
    title = "{Nonperturbative evolution equation for quantum gravity}",
    eprint = "hep-th/9605030",
    archivePrefix = "arXiv",
    reportNumber = "DESY-96-080",
    doi = "10.1103/PhysRevD.57.971",
    journal = "Phys. Rev. D",
    volume = "57",
    pages = "971--985",
    year = "1998"
}

@article{Ricon19,
    author = "Contreras, Ernesto and Rinc{\'o}n, {\'A}ngel and Panotopoulos, Grigoris and Bargue{\~n}o, Pedro and Koch, Benjamin",
    title = "{Black hole shadow of a rotating scale--dependent black hole}",
    eprint = "1906.06990",
    archivePrefix = "arXiv",
    primaryClass = "gr-qc",
    doi = "10.1103/PhysRevD.101.064053",
    journal = "Phys. Rev. D",
    volume = "101",
    number = "6",
    pages = "064053",
    year = "2020"
}

@article{Rincon2020,
    author = "Rinc{\'o}n, {\'A}ngel and Panotopoulos, Grigoris",
    title = "{Quasinormal modes of an improved Schwarzschild black hole}",
    eprint = "2006.11889",
    archivePrefix = "arXiv",
    primaryClass = "gr-qc",
    doi = "10.1016/j.dark.2020.100639",
    journal = "Phys. Dark Univ.",
    volume = "30",
    pages = "100639",
    year = "2020"
}

@article{RovelliPS,
    author = "Rovelli, Carlo and Vidotto, Francesca",
    title = "{Planck stars}",
    eprint = "1401.6562",
    archivePrefix = "arXiv",
    primaryClass = "gr-qc",
    doi = "10.1142/S0218271814420267",
    journal = "Int. J. Mod. Phys. D",
    volume = "23",
    number = "12",
    pages = "1442026",
    year = "2014"
}

@article{Sagnotti1985,
    author = "Goroff, Marc H. and Sagnotti, Augusto",
    title = "{The Ultraviolet Behavior of Einstein Gravity}",
    reportNumber = "CALT-68-1289, LBL-19995, UCB-PTH-85-34",
    doi = "10.1016/0550-3213(86)90193-8",
    journal = "Nucl. Phys. B",
    volume = "266",
    pages = "709--736",
    year = "1986"
}

@article{Sasaki2025,
    author = "M. Sasaki and J. Wang",
    title = "{Unveiling Primordial Black Hole Relics Through Induced Gravitational Waves}",
    eprint = "2512.22450",
    archivePrefix = "arXiv",
    primaryClass = "gr-qc",
    journal = "arXiv",
    year = "2025"
}

@article{Scardigli2010,
    author = "Scardigli, Fabio and Gruber, Christine and Chen, Pisin",
    title = "{Black Hole Remnants in the Early Universe}",
    eprint = "1009.0882",
    archivePrefix = "arXiv",
    primaryClass = "gr-qc",
    doi = "10.1103/PhysRevD.83.063507",
    journal = "Phys. Rev. D",
    volume = "83",
    pages = "063507",
    year = "2011"
}

@article{Scardigli2017PLB,
    author = "Scardigli, Fabio and Lambiase, Gaetano and Vagenas, Elias",
    title = "{GUP parameter from quantum corrections to the Newtonian potential}",
    eprint = "1611.01469",
    archivePrefix = "arXiv",
    primaryClass = "hep-th",
    doi = "10.1016/j.physletb.2017.01.054",
    journal = "Phys. Lett. B",
    volume = "767",
    pages = "242--246",
    year = "2017"
}

@ARTICLE{Scardigli2023PRD,
  author = {F. Scardigli and G. Lambiase},
  title = {Planck Stars from a Scale-dependent Gravity theory},
  journal = {Phys. Rev. D},
  volume = {107},
  pages = {104001},
  year = {2023}
}

@article{Shapiro2022,
    author = "de Paula Netto, Tib{\'e}rio and Modesto, Leonardo and Shapiro, Ilya L.",
    title = "{Universal leading quantum correction to the Newton potential}",
    eprint = "2110.14263",
    archivePrefix = "arXiv",
    primaryClass = "hep-th",
    doi = "10.1140/epjc/s10052-022-10077-7",
    journal = "Eur. Phys. J. C",
    volume = "82",
    number = "2",
    pages = "160",
    year = "2022"
}

@inproceedings{Spina24,
    author = "Spina, Andrea and Silveravalle, Samuele and Bonanno, Alfio",
    title = "{Scalar Perturbations of Regular Black Holes derived from a Non-Singular Collapse Model in Asymptotic Safety}",
    booktitle = "{17th Marcel Grossmann Meeting}",
    eprint = "2410.05936",
    archivePrefix = "arXiv",
    primaryClass = "gr-qc",
    month = "10",
    year = "2024"
}

@article{Stashko2024PRD,
title={Quasinormal modes and gray-body factors of regular black holes in asymptotically safe gravity},
author={O. Stashko},
  journal={Phys. Rev. D},
  volume={110},
  pages={084016},
  year={2024}
}

@article{Stelle1977,
    author = "Stelle, K. S.",
    title = "{Renormalization of Higher Derivative Quantum Gravity}",
    reportNumber = "PRINT-76-1059 (BRANDEIS)",
    doi = "10.1103/PhysRevD.16.953",
    journal = "Phys. Rev. D",
    volume = "16",
    pages = "953--969",
    year = "1977"
}

@article{Teukolsky1972,
    author = "Teukolsky, S. A.",
    title = "{Rotating black holes - separable wave equations for gravitational and electromagnetic perturbations}",
    reportNumber = "OAP-291",
    doi = "10.1103/PhysRevLett.29.1114",
    journal = "Phys. Rev. Lett.",
    volume = "29",
    pages = "1114--1118",
    year = "1972"
}

@article{Verlinde2011,
    author = "Verlinde, Erik P.",
    title = "{On the Origin of Gravity and the Laws of Newton}",
    eprint = "1001.0785",
    archivePrefix = "arXiv",
    primaryClass = "hep-th",
    doi = "10.1007/JHEP04(2011)029",
    journal = "JHEP",
    volume = "04",
    pages = "029",
    year = "2011"
}

@inproceedings{Weinberg1,
    author = "Weinberg, Steven",
    title = "{Critical Phenomena for Field Theorists}",
    booktitle = "{14th International School of Subnuclear Physics: Understanding the Fundamental Constitutents of Matter}",
    reportNumber = "HUTP-76-160",
    doi = "10.1007/978-1-4684-0931-4_1",
    month = "8",
    year = "1976"
}

@inbook{Weinberg2,
    author = "Weinberg, Steven",
    title = "{Ultaviolet divergences in quantum theory of gravitation}",
    booktitle = "{General Relativity}: {An Einstein Centenary Survey}",
    pages = "790--831",
    year = "1980"
}

@article{Wetterich93,
    author = "Wetterich, Christof",
    title = "{Exact evolution equation for the effective potential}",
    eprint = "1710.05815",
    archivePrefix = "arXiv",
    primaryClass = "hep-th",
    reportNumber = "HD-THEP-92-61",
    doi = "10.1016/0370-2693(93)90726-X",
    journal = "Phys. Lett. B",
    volume = "301",
    pages = "90--94",
    year = "1993"
}

@article{Yi2024,
    author = "Yi, Sophia and Kuntz, Adrien and Barausse, Enrico and Berti, Emanuele and Cheung, Mark Ho-Yeuk and Kritos, Konstantinos and Maselli, Andrea",
    title = "{Nonlinear quasinormal mode detectability with next-generation gravitational wave detectors}",
    eprint = "2403.09767",
    archivePrefix = "arXiv",
    primaryClass = "gr-qc",
    reportNumber = "ET-0081A-24",
    doi = "10.1103/PhysRevD.109.124029",
    journal = "Phys. Rev. D",
    volume = "109",
    number = "12",
    pages = "124029",
    year = "2024"
}

@inproceedings{Zinhailo2023,
    author = "Zinhailo, A. F.",
    title = "{Quasinormal spectrum in the asymptotically safe gravity}",
    eprint = "2311.05380",
    archivePrefix = "arXiv",
    primaryClass = "gr-qc",
    month = "11",
    year = "2023"
}

@book{Ince1956,
    author ={E. L. Ince},
    title ={Ordinary Differential Equations},
    publisher = {Dover: New York},
    year = {1956}
}

@book{Boyd2000,
   author = {J. P. Boyd},
   city = {New York},
   edition = {2nd},
   pages = {688},
   publisher = {Dover Publications, New York},
   title = {Chebyshev and Fourier Spectral Methods},
   year = {2000},
}

@book{Trefethen2000,
   author = {L. N. Trefethen},
   pages = {184},
   publisher = {Society for Industrial and Applied Mathematics, Philadelphia, PA, USA},
   title = {Spectral methods in MatLab},
   url = {http://web.comlab.ox.ac.uk/oucl/work/nick.trefethen/spectral.html},
   year = {2000},
}

@book{Fox1968,
   author = {L. Fox and I. B. Parker},
   city = {Oxford},
   pages = {205},
   publisher = {Oxford University Press},
   title = {Chebyshev Polynomials in Numerical Analysis},
   year = {1968},
}

@book{mct2015,
 author = {P. Holodoborodko},
 title = {{Multiprecision Computing Toolbox for MATLAB 5.2.7.15522}},
 publisher = {Advanpix LLC.},
 address = {Yokohama, Japan},
 date = {2023-11-29},
 year = {2023}
}

@article{Tisseur2001,
   author = {F. Tisseur and K. Meerbergen},
   doi = {10.1137/S0036144500381988},
   issn = {0036-1445},
   issue = {2},
   journal = {SIAM Review},
   month = {1},
   pages = {235-286},
   title = {The Quadratic Eigenvalue Problem},
   volume = {43},
   year = {2001},
}
\end{document}